\documentclass[fleqn,usenatbib]{mnras}

\usepackage{newtxtext,newtxmath}

\usepackage[T1]{fontenc}
\usepackage{ae,aecompl}

\usepackage{graphicx}	
\usepackage{amsmath}	
\usepackage{amssymb}	

\usepackage{todonotes}
\usepackage[normalem]{ulem}

\usepackage[cal=boondox]{mathalfa}

\usepackage{array}
\usepackage{booktabs}
\usepackage{color}
\usepackage{subfig}
\usepackage{verbatim}
\usepackage{bm}

\usepackage{tikz}
\usetikzlibrary{shapes,arrows}

\usepackage{times}

\newcommand{\hinv}{h^{-1}}
\newcommand{\beq}{\begin{equation}}
\newcommand{\eeq}{\end{equation}}

\newcommand{\nhat}{\hat{\textbf{n}}}
\newcommand{\dint}{\mathrm{d}}

\newcommand{\mb}{\mathbf}

\newcommand{\Planck}{{\slshape Planck~}}

\newcommand{\arcm}{\mathrm{arcmin}}
\newcommand{\msol}{h^{-1}M_{\odot}}
\newcommand{\Msol}{M_{\odot}}
\newcommand{\mpc}{h^{-1}\mathrm{Mpc}}

\newcommand{\Mpc}{\mathrm{Mpc}}

\newcommand{\ellta}{\ell \theta_A}

\title[Pairwise kSZ for cosmology]{Cosmology with the pairwise kinematic SZ effect: 
Calibration and validation using hydrodynamical simulations
}

\author[B. Soergel et al.]{
	\parbox{\textwidth}{Bjoern Soergel$^{1,}$\thanks{E-mail: bsoergel@ast.cam.ac.uk},
		Alexandro Saro$^{2,3}$,
		Tommaso Giannantonio$^{2,1}$,
		George Efstathiou$^{1}$,
		and Klaus Dolag$^{2,4}$
		\vspace{2mm}
	}
	\\
	$^{1}$ Institute of Astronomy \& Kavli Institute for Cosmology,  University of Cambridge, Madingley Road, Cambridge CB3 0HA, UK\\
	$^{2}$ Universit\"{a}ts-Sternwarte, Ludwig-Maximilians-Universit\"{a}t M\"{u}nchen, Scheinerstr. 1, 81679 M\"{u}nchen, Germany\\
	$^{3}$ INAF-Osservatorio Astronomico di Trieste, via G.B. Tiepolo 11, 34131, Trieste, Italy\\
	$^{4}$Max-Planck-Institut f\"{u}r Astrophysik, Karl-Schwarzschild Strasse 1, 85748 Garching, Germany
}

\pubyear{2018}

\begin{document}
\label{firstpage}
\pagerange{\pageref{firstpage}--\pageref{lastpage}}
\maketitle

\begin{abstract}
We study the potential of the kinematic SZ effect as a probe for cosmology, focusing on the pairwise method. The main challenge is disentangling the cosmologically interesting mean pairwise velocity from the cluster optical depth and the associated uncertainties on the baryonic physics in clusters. Furthermore, the pairwise kSZ signal might be affected by internal cluster motions or correlations between velocity and optical depth. We investigate these effects using the \textit{Magneticum} cosmological hydrodynamical simulations, one of the largest simulations of this kind performed to date.
We produce tSZ and kSZ maps with an area of $\simeq 1600~\mathrm{deg}^2$, and the corresponding cluster catalogues with $M_{500c} \gtrsim 3 \times 10^{13}~h^{-1}M_\odot$ and $z \lesssim 2$.
From these data sets we calibrate a scaling relation between the average Compton-$y$ parameter and optical depth. We show that this relation can be used to recover an accurate estimate of the mean pairwise velocity from the kSZ effect, and that this effect can be used as an important probe of cosmology. 
We discuss the impact of theoretical and observational systematic effects, and find that further work on feedback models is required
to interpret future high-precision measurements of the kSZ effect.
\end{abstract}




\section{Introduction}
\label{sec:intro}

When passing through a cluster of galaxies, a small fraction of cosmic microwave background (CMB) photons are scattered off electrons in the hot, ionized intra-cluster medium (ICM). 
This process, known as the Sunyaev-Zel'dovich (SZ) effect, can give rise to both a change in the CMB blackbody temperature as well as a spectral distortion (\citealt{SZ1970,SZ1972,SZ1980}; see also \citealt{Birkinshaw1998,Carlstrom2002} for reviews).
While the underlying physical process, the scattering of CMB photons off moving electrons, is the same, the SZ signal is usually broken down into two separate contributions: the \textit{thermal} and \textit{kinematic} SZ effects.

The \textit{thermal} SZ (tSZ) signal is caused by the high random velocities of the electrons in the hot intra-cluster medium, giving rise to a characteristic spectral distortion to the CMB blackbody. Over the past decade, the tSZ effect has become a well-established tool for blind cluster finding in CMB surveys (e.g.~\citealt{bleem15,Hasselfield2013,Planck2015_SZ}), as well as for determining the thermal properties of the ICM in stacked observations (e.g.~\citealt{Afshordi07,Diego10,plagge10,Bartlett11,Planck12,Planckint_hotgas,Hajian13,Soergel2017,Saro2017,Erler2017}).

On the other hand, the \textit{kinematic} SZ (kSZ) effect is a result of the bulk motion of the entire cluster, leading to a Doppler shift in the CMB temperature while preserving its blackbody spectrum.
This signal offers significant potential for constraints on both astrophysics and cosmology (e.g.~\citealt{Rephaeli1991,Haehnelt1996,Diaferio2000,Aghanim2001,Bhattacharya2006}).
Because of its small amplitude 
(the ratio of kSZ to tSZ is $ \mathrm{v}/c \times (k_\mathrm{B}T/m_ec^2)^{-1} \simeq 0.1$ for typical clusters) 
and the lack of a distinctive frequency signature, the kSZ effect has been significantly harder to detect than its thermal counterpart. 
Nonetheless, it can be extracted using a differential statistic probing the mean relative momentum of the cluster sample \citep{Hand2012}.
Applying this \textit{pairwise} estimator to CMB data from the Atacama Cosmology Telescope (ACT, \citealt{ACT}) and spectroscopic galaxy data from the Baryon Oscillation Spectroscopic Survey (BOSS, \citealt{Ahn2012}), \cite{Hand2012} obtained the first detection of the kSZ effect.

Subsequently, \citet{Planck_KSZ} reported a detection obtained with the same estimator. 
\citet{Soergel2016} demonstrated that this technique can also be used with photometric galaxy cluster catalogues, leading to a detection using data from the Dark Energy Survey (DES, \citealt{DES,DESnonDEoverview}) and the South Pole Telescope (SPT, \citealt{lueker10}).
In \cite{deBernardis2016}, the original analysis by \cite{Hand2012} was updated to include more recent ACT and BOSS data; they also detected the expected imprint of redshift-space distortions on the kSZ signal \citep{Sugiyama2016}. 
Recently \cite{Calafut2017} and \cite{LiYC2017} reported further measurements (albeit at low-significance) by combining \Planck data with various SDSS-based galaxy and cluster samples, and BOSS data, respectively.
Furthermore, \cite{Sugiyama2017} obtained a detection from \Planck and BOSS data using a Fourier-space version of the pairwise estimator.

There are also alternative statistical approaches: on the one hand it is possible to correlate a CMB map with a reconstructed velocity field \citep{Li2014}; this has led to detections from both \Planck and ACT data (\citealt{Planck_KSZ} and \citealt{Schaan2015}, respectively). 
The measurements of the \Planck analysis were used by \cite{Ma2017} to simultaneously constrain optical depth and velocity bias of the used galaxy sample.
\citet{Hill2016} and \citet{Ferraro2016} measured the kSZ imprint on the cross-correlation of a squared CMB map with a tracer of the projected density field.
\cite{Planck_kSZ_dispersion} reported a detection based on an increased variance in the observed CMB temperature at the positions of clusters.
Finally, there have also been kSZ measurements with high-resolution multi-frequency observations of individual clusters (e.g.~\citealt{Sayers2013}). 

All these analyses, both with the pairwise and alternative techniques, are close to the detection threshold,
with reported significances in the $\simeq 2-4\sigma$ range. 
However, with ongoing and upcoming CMB and large-scale structure surveys, detection significances will increase dramatically (e.g.~\citealt{Keisler2012,Flender2015,Alonso2016,Ferraro2016}).
Therefore the kSZ effect is expected to mature rapidly into a cosmological probe.
In this paper, we test whether the kSZ signal can be used as an accurate probe of cosmology.
We focus on the pairwise statistic because it is a robust detection method with a well-studied astrophysical and cosmological sensitivity.
In particular, previous works (\citealt{Soergel2016,deBernardis2016,Flender2017}) have assumed the simple relation
\beq
T_\mathrm{pkSZ}(r,z) \simeq \overline{\tau} \frac{T_\mathrm{CMB}}{c} \mathcal{v}_{12}(r,z)
\label{eq:pksz}
\eeq
between the pairwise kSZ signal $T_\mathrm{pkSZ}$ and the average optical depth $\overline{\tau}$ and mean pairwise velocity $\mathcal{v}_{12}$ of the cluster sample.
In a fixed cosmology, and given a prescription for the halo bias, $\mathcal{v}_{12}(r,z)$ can be calculated (e.g.~\citealt{Sheth2000}).
Therefore $\overline{\tau}$ can be determined with a simple template fit;
its variation with aperture or filtering scale is then a probe of the gas fraction and density profile of the clusters \citep{HM2015,Soergel2016,Sugiyama2017}.

If, on the other hand, we want to obtain cosmological constraints from the pairwise kSZ measurement, this argument needs to be turned around: 
we require a \textit{prediction} for the optical depth to reconstruct an estimate of the mean pairwise velocity (\citealt{Dolag2013,Battaglia2016,Flender2017}). 
This measurement then constrains the parameter combination $Hf\sigma_8^2$ \citep{Keisler2012,Sugiyama2017}, where $H$ is the Hubble parameter, $f$ is the linear growth rate of density perturbations, and $\sigma_8$ is their amplitude averaged over a sphere of $8~\hinv\Mpc$.\footnote{When redshift-space effects are included, there are additional corrections, and the pairwise kSZ signal needs to be decomposed into multipoles \citep{Sugiyama2017}.} 
Therefore the pairwise kSZ signal can be used to constrain various aspects of cosmology, including dark energy or modified gravity \citep{Bhattacharya2006,Bhattacharya2007,Keisler2012,Mueller2014a} or massive neutrinos \citep{Mueller2014b}.

In this paper we test whether the assumption of eq.~\ref{eq:pksz} in fact holds at the precision required to extract cosmology from the high-significance pairwise kSZ measurements that are expected from future surveys.
Internal cluster motions, bulk rotation, or relative velocities between dark matter and gas can introduce significant scatter in the relation between halo peculiar velocity and observed kSZ signal \citep{Cooray2002,Chluba2002,Nagai2003,Diaferio2004}.
These effects could also cause a bias to the observed pairwise kSZ signal \citep{Dolag2013}, for example if internal motions are correlated with the environment or bulk motion of a cluster.
The approximation of eq.~\ref{eq:pksz} could also break down if there is a non-negligible degree of correlation between velocity and optical depth of haloes.
We show how these two potential biases propagate into the observed pairwise kSZ signal in Sec.~\ref{subsec:pairwise} below.

Modelling these effects requires hydrodynamical simulations with sufficient resolution to resolve internal cluster dynamics.
Furthermore, simulations of the pairwise kSZ signal require a large box size to capture the large-scale motions of clusters on $\gtrsim 100$~Mpc scales.
The combination of these two requirements has made these simulations prohibitively expensive until recently.
In this work, we use the \textit{Magneticum} simulation suite\footnote{\url{http://www.magneticum.org/}}, which contains amongst others one of the largest hydrodynamical simulations performed to date. 
This enables us to create simulated cluster catalogues and tSZ and kSZ maps that are well matched to upcoming CMB and large-scale structure surveys, both in terms of sky coverage and mass and redshift range.
With these, we are able to test the potential of the pairwise kSZ signal as a cosmological probe in detail.

Our paper is structured as follows: In Sec.~\ref{sec:theory_sims} we briefly review the theory of both SZ effects and describe the simulation and the production of cluster catalogues and SZ maps.
Our analysis methods for predicting the optical depth and measuring the pairwise kSZ signal are detailed in Sec.~\ref{sec:methods}.
In Sec.~\ref{sec:results} we show the resulting measurement of the mean pairwise velocity and compare it to the true value obtained directly from the halo catalogue.
We further discuss the implications for the pairwise kSZ signal as a cosmological probe, before concluding in Sec.~\ref{sec:conclusions}.

\section{Theory and simulations}
\label{sec:theory_sims}
\subsection{SZ effects}
\label{subsec:SZ}

\subsubsection{Kinematic SZ}
The kSZ effect preserves the blackbody spectrum of the CMB, but imparts a Doppler shift given by \citep{SZ1980}
\beq
\frac{\Delta T_\mathrm{kSZ}}{T_{\mathrm{CMB}}} = -  \sigma_T \int \dint l \,  n_{e}(\mathbf{r}) \frac{\hat{\mathbf{r}} \cdot \mathbf{v}_e(\mathbf{r})}{c}
\simeq - \tau \frac{\mathrm{v}_\mathrm{los}}{c} \, ,
\label{eq:ksz}
\eeq
where $n_e$ and $\mathbf{v}_e$ are the number density and peculiar velocity of electrons in a cluster at direction $\hat{\mathbf{r}}$ and $\sigma_T$ is the Thomson cross section.
If internal motions of the cluster gas are negligible (which might not be a good approximation; see above), the kSZ signal reduces to the product of cluster line-of-sight velocity and the optical depth
\beq
\tau  = \int  \mathrm{d}l\,  n_{e}(\mathbf{r}) \sigma_T \, .
\label{eq:tau}
\eeq

\subsubsection{Thermal SZ}
The tSZ effect, on the other hand, induces a spectral distortion of the microwave background.
In the non-relativistic limit it is given by \citep{SZ1970}
\beq
\frac{\Delta T_\mathrm{tSZ}}{T_{\mathrm{CMB}}} = y f(x) \quad \text{with} \quad f(x) = x \coth(x/2) - 4 \, .
\eeq
Here  $x = h\nu / (k_\mathrm{B}T_\mathrm{CMB})$ is a dimensionless frequency, and the Compton-$y$ parameter is given by
\beq
y=\int \mathrm{d}l \, n_e(\mb{r}) \, \sigma_T \, \frac{k_\mathrm{B} T_e(\mb{r})}{m_e c^2}    \, .
\label{eq:tsz}
\eeq
While relativistic corrections are non-negligible for massive clusters with electron temperature $k_\mathrm{B} T_e \gtrsim 10~\mathrm{keV}$ (e.g.~\citealt{Carlstrom2002}), these objects are rare.
The bulk of the pairwise kSZ signal is coming from much lower masses, so neglecting relativistic corrections to the tSZ signal does not introduce a significant bias into our analysis.

\subsection{Magneticum simulations}
\label{subsec:sims}

\begin{figure*}
	\includegraphics[width=\columnwidth]{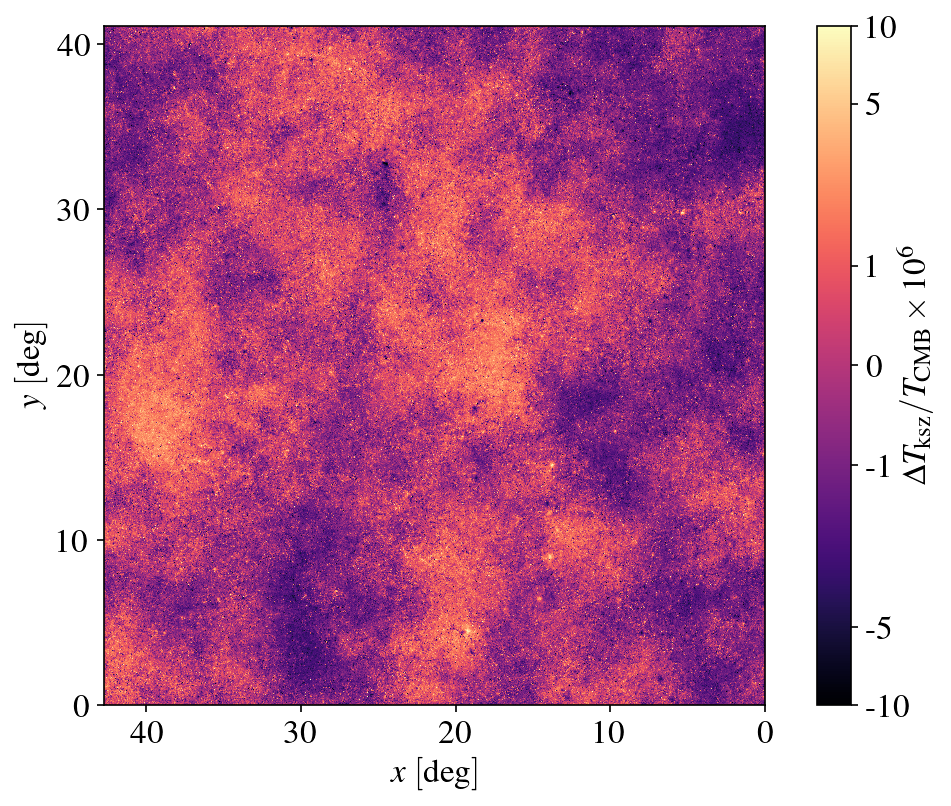}
	\includegraphics[width=\columnwidth]{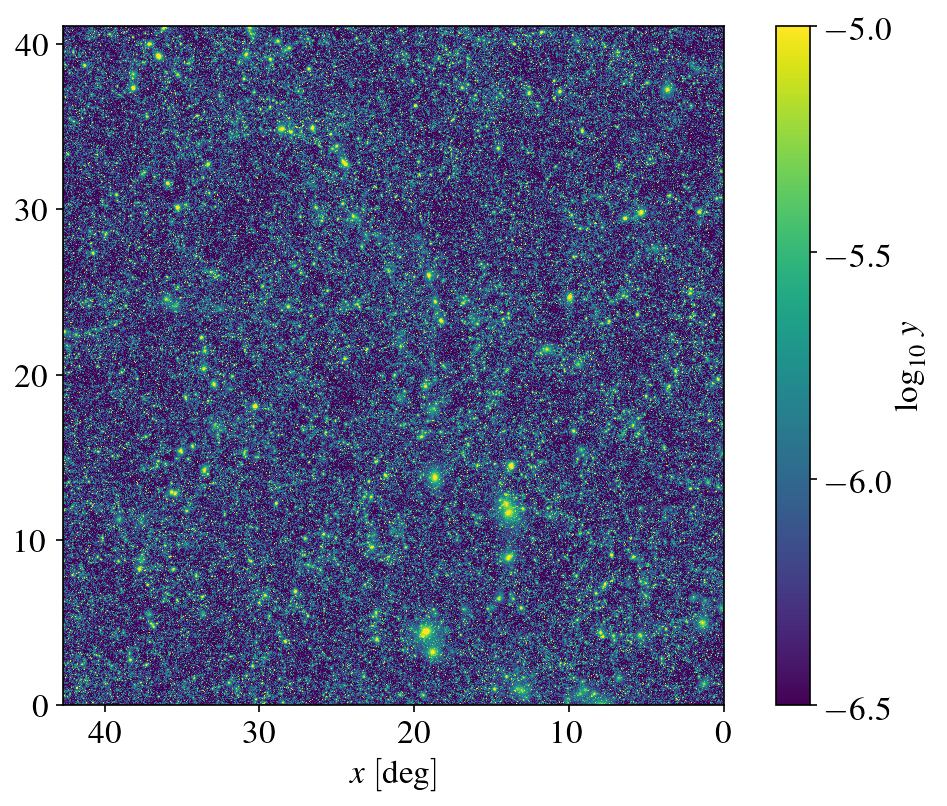}
	\caption{SZ maps from the \textit{Magneticum} simulations: We show on the left the kSZ signal, where we have used an arcsinh(..) colour mapping to increase the dynamic range, while at the same time preserving the sign of the kSZ. For the tSZ signal in the right panel we show the logarithm of the Compton-$y$ parameter. 
	}
	\label{fig:ksz_tsz_map}
\end{figure*}

Here we only describe the simulations briefly; for a more detailed description we refer to previous work using these simulations (e.g.~\citealt{Dolag2015,Bocquet2016,Gupta2016}).
The {\it Magneticum} simulations are a set of state-of-the-art, cosmological hydrodynamical simulations of
different cosmological volumes with different resolutions performed with an improved version of \texttt{GADGET3} \citep{Springel2005,2016MNRAS.455.2110B}. 
They follow a standard $\Lambda$CDM cosmology with
parameters close to the best-fitting values from WMAP7 \citep{Komatsu2011}, i.e.~a matter density $\Omega_m = 0.272$,
baryon density $\Omega_b = 0.046$ and a Hubble constant of $H_0 = 70.4~\mathrm{km}~\mathrm{s}^{-1}~\mathrm{Mpc}^{-1}$.
The spectral index of the primordial power spectrum is $n_s = 0.963$, whereas the normalization of the present-day power
spectrum is given by $\sigma_8 = 0.809$.

These simulations follow a wide range of physical processes (see \citealt{Hirschmann2014,Teklu2015} for details)
which are important for studying the formation of active galactic nuclei (AGN), galaxies, and galaxy clusters.
The simulation set covers a huge dynamical range following the same underlying treatment of the
physical processes controlling galaxy formation, thereby allowing to reproduce the properties of
the large-scale, intra-galactic and intra-cluster medium 
\citep{Dolag2015,Gupta2016,Remus2017b},
as well as the detailed properties of galaxies including morphological classifications and internal properties 
\citep{Teklu2015,Remus2017,Teklu2017}.
This also includes the distribution of different metal species within galaxies and galaxy clusters \citep{2017Galax...5...35D}, and the properties of
the AGN population within the simulations \citep{Hirschmann2014,Steinborn2016}.

Here we focus on the largest box (`Box0') with size \mbox{$L = 2688~\mpc$} and $2 \times 4536^3$ particles
\citep[see also][]{Bocquet2016,Pollina2017}, making it the largest cosmological hydrodynamical simulation
performed to date. Such a large box size is essential for our purposes as the pairwise kSZ signal only
vanishes at scales $r \gtrsim 300~\Mpc$; furthermore large cluster samples are required to analyse the kSZ signal.

We identify halos using a friends-of-friends algorithm with linking length $b=0.16$. For the identified halos we
then compute their properties -- like spherical overdensity masses -- using the \texttt{SUBFIND} algorithm
\citep{Springel2001,Dolag2009}; 
see \cite{Bocquet2016} and the discussion therein for the obtained halo mass function.
We note that even the smallest halos we consider ($M_{500c} \gtrsim 3 \times 10^{13}~\msol$) are still resolved
by $\gtrsim 2000$ dark matter particles and the associated number of gas and star particles.

\subsection{SZ maps}
\label{subsec:maps}

\begin{table}
	\begin{tabular}{ccccc}
\toprule
Redshift slice & $z_\mathrm{tab}$ & width [cMpc] & depth [cMpc] \\
\midrule
$0.00 < z < 0.07$ &            0.001 &          101 &          288 \\
$0.07 < z < 0.21$ &            0.138 &          406 &          585 \\
$0.21 < z < 0.38$ &            0.294 &          826 &          615 \\
$0.38 < z < 0.57$ &            0.472 &         1264 &          636 \\
$0.57 < z < 0.78$ &            0.673 &         1713 &          645 \\
$0.78 < z < 1.04$ &            0.903 &         2171 &          664 \\
$1.04 < z < 1.32$ &            1.181 &         2631 &          652 \\
$1.32 < z < 1.59$ &            1.480 &         3042 &          521 \\
$1.59 < z < 1.84$ &            1.709 &         3375 &          429 \\
$1.84 < z < 2.15$ &            1.983 &         3689 &          469 \\
\bottomrule
\end{tabular}

	\caption{Redshift slices used for the creation of the lightcone.}
	\label{tab:geo}
\end{table}

Ideally, one would create the SZ maps by placing an observer at one corner of the simulation box. One would then solve the lightcone equation for every particle, interpolating their positions between the snapshots (as e.g.~in~\citealt{Flender2015}).
Only gas particles within the lightcone would then contribute to the SZ map.
This approach is, however, not feasible for hydrodynamical simulations, as the gas properties cannot be interpolated safely between snapshots (one would miss shocks, for example). 
Furthermore, this problem will be exacerbated if the stored snapshots are relatively far apart, as it is necessarily the case for such a large simulation as the one considered here.
Producing the maps on the fly without relying on snapshots, on the other hand, would add significantly to the computational complexity of the simulation.

We therefore approximate the full lightcone by a series of redshift slices with $\Delta z \simeq 0.2$, making use of the \texttt{SMAC} code \citep{Dolag2005}.\footnote{\url{http://wwwmpa.mpa-garching.mpg.de/~kdolag/Smac/}}
For every slice, we take a region of appropriate width and depth from the simulation box at the corresponding redshift $z_\mathrm{tab}$ (see Table~\ref{tab:geo}).
We then calculate the kSZ and tSZ signal from the slices via eqs.~\ref{eq:ksz} and~\ref{eq:tsz} by projecting along the $z$-axis of the simulation box.
The maps for the full lightcone are then simply given by the co-addition of the signals from the individual slices.
\footnote{
We have made the simulated SZ maps and the cluster lightcone data available at \url{http://magneticum.org/data.html\#SZ}.
}
We show the final kSZ and tSZ maps in Fig.~\ref{fig:ksz_tsz_map}.

The map size in this approach is determined by the size of the simulation box and the highest desired redshift.
For $z_\mathrm{max} \simeq 2$, this yields a map size of $\simeq 40~\deg$, corresponding to an area of $\simeq 1600~\deg^2$.\footnote{Larger sky coverages can, of course, be achieved if $z_\mathrm{max}$ is reduced. In addition, one could project the box along different axes; the duplication of structures caused by this approach would be minimal.}
We note that this is sufficiently large to match the sky coverage of current and future analyses correlating high-resolution CMB and large-scale structure data.
For reference, the effective overlapping sky area between DES Year 1 and SPT is around 1,200 deg$^2$ \citep{Soergel2016}.

\begin{figure}
	\includegraphics[width=\columnwidth]{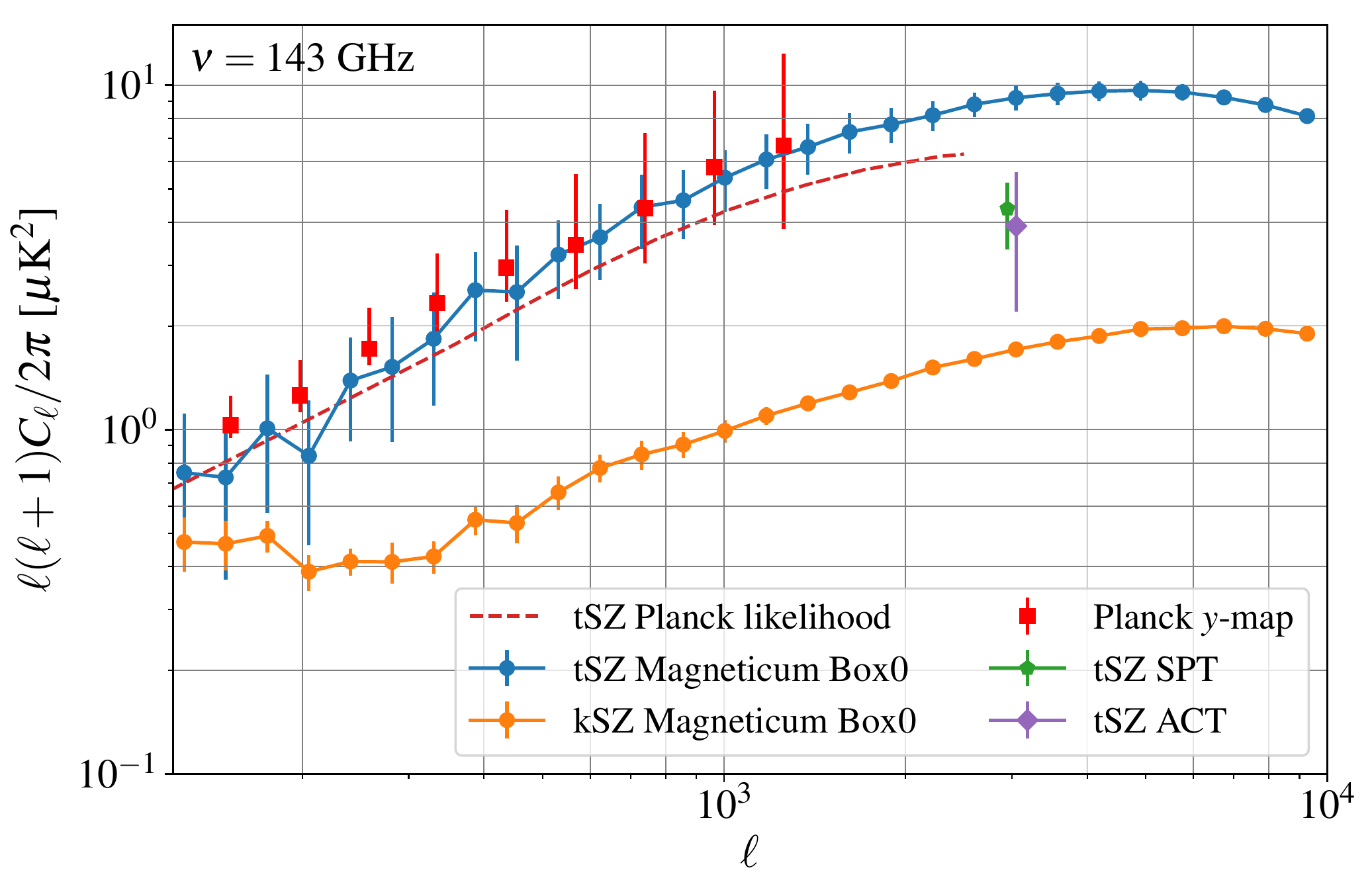}
	\caption{
	SZ power spectra: We show here the kSZ and tSZ power spectra from the \textit{Magneticum} simulation, where the error bars are estimated by splitting the map into $5 \times 5$ rectangular sub-patches.
	For the tSZ component, we compare to the power spectrum of the \Planck $y$-map \citep{Planck2015_SZmap}, the best-fitting tSZ template from the \Planck likelihood \citep{Planck2015params}, and to measurements of the tSZ power at $\ell=3000$
	from ACT \citep{Dunkley2013} and SPT \citep{George2014}.
	It is worth noting that the latter two measurements used tSZ templates that differ in shape from our simulation results. This will likely have an impact on the level of agreement between ACT/SPT and the simulated SZ power spectrum (see also Fig.~6 in \citealt{Dolag2015}).
	In all cases, we have scaled the measurements to a frequency of 143 GHz.
	}
	\label{fig:Cls}
\end{figure}

As part of the validation of our map pipeline, we compute the power spectrum of the kSZ and tSZ map in the range \mbox{$\ell = [100,10\,000]$} and show them in Fig.~\ref{fig:Cls}.
\cite{Dolag2015} have previously compared SZ power spectra from the second-largest \textit{Magneticum} box (`Box1', $L = 896~\mpc$) to observations.
They found the predicted tSZ power spectrum to be in good agreement with \Planck ($\ell \lesssim 1000$), but higher than measured by ACT \citep{Dunkley2013} and SPT \citep{George2014} on smaller scales.
Here we observe the same for the largest \textit{Magneticum} box.
The kSZ power spectrum at multipoles below a few hundred qualitatively matches with the contribution expected from longitudinal momentum modes at large scales (e.g.~\citealt{Park2015}). At smaller scales only perpendicular modes contribute significantly to the kSZ power spectrum (e.g.~\citealt{Jaffe1998}).

It is worth noting that especially the high-$\ell$ tSZ power spectrum is sensitive to the sub-grid physics of the simulation, and particularly to the strength of AGN feedback (e.g.~\citealt{McCarthy2014}).
The same holds true for the kSZ power spectrum; see e.g.~\citet{Park2017} for a recent study using the Illustris simulation.
Given that we are focussing on a single sub-grid model for the feedback processes here, there is inevitably a systematic uncertainty associated with our predictions; see Sec.~\ref{subsec:systematics} below for a further discussion.

\subsection{Mass and redshift selection}

\begin{figure}
	\includegraphics[width=\columnwidth]{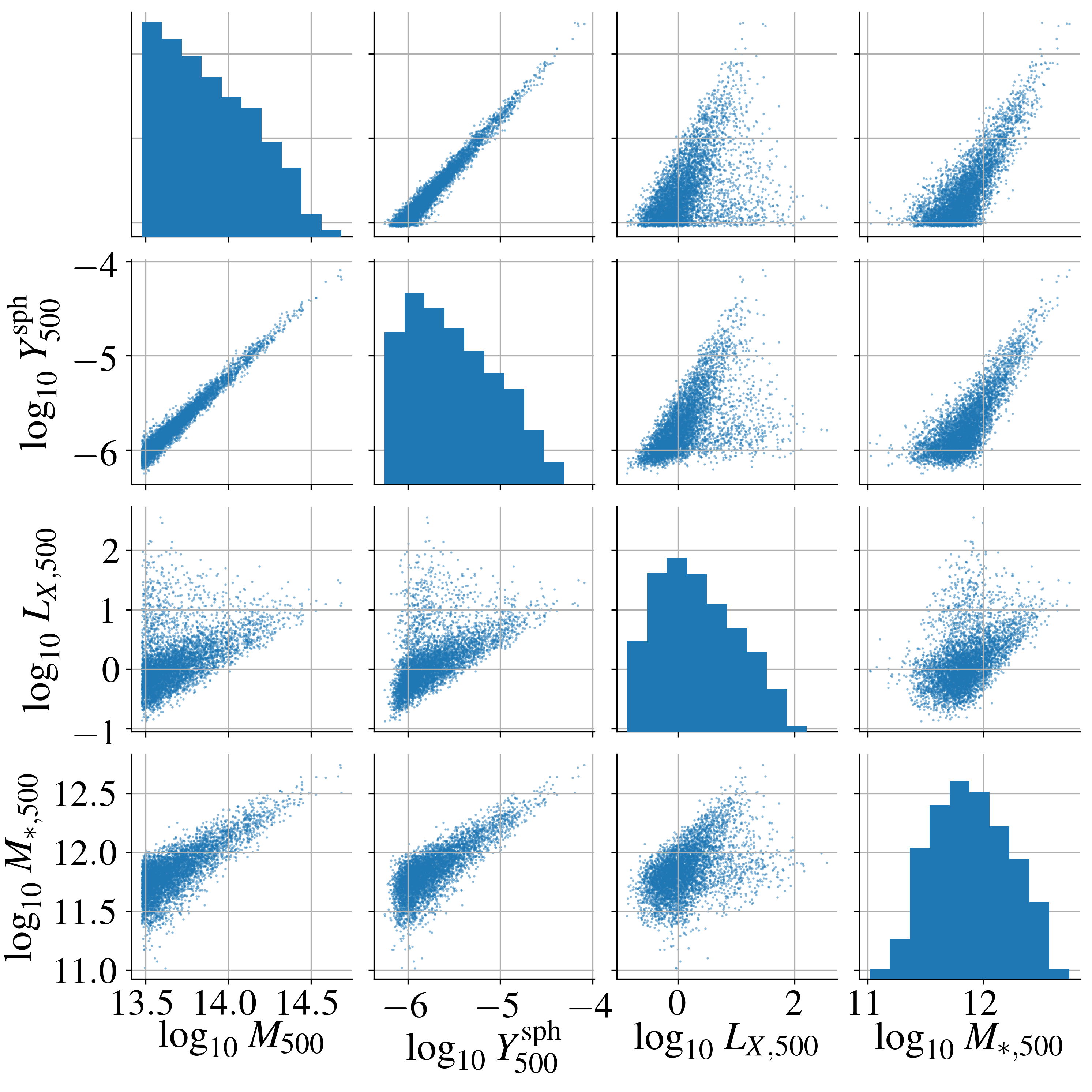}
	\caption{Selection of the cluster properties calculated during the simulation post-processing: On the diagonals we show the distribution of
	 $M_{500}~[\msol]$, the integrated Compton-$y$ parameter $Y_{500}~[\mathrm{Mpc}^2]$, X-ray luminosity $L_{X,500}~[10^{44}~\mathrm{erg}/\mathrm{s}]$ and stellar mass $M_{\ast,500}~[\msol]$. 
	 The off-diagonal panels show the correlations between the respective quantities. The lower mass limit was set to $M_{500c} > 3 \times 10^{13}~\msol$.
	}
	\label{fig:bigscatter}
\end{figure}

For the main analysis we select clusters by imposing a lower mass limit of $M_{500c} > 3 \times 10^{13}~\msol$,
leaving around 65,000 clusters in our sample.
During the post-processing of the simulation data we have also calculated a variety of other observationally relevant quantities, such as X-ray luminosity, stellar mass, and integrated SZ signal.
The distribution of these quantities (all within $R_{500c}$), and how they correlate with each other, is shown in Fig.~\ref{fig:bigscatter}.
It would also be possible to select clusters by various mass proxies, such as $L_X$, $M_\ast$, or $Y_\mathrm{SZ}$.
Therefore we can, in principle, create mock catalogues for clusters selected in the X-ray, optical or via their SZ signature;
in this case, the scatter in the mass-observable relations would cause a Malmquist-type bias in the selected catalogue (see Fig.~\ref{fig:bigscatter}).
A detailed mock-up of observational cluster selection with all its subtleties is, however, beyond the scope of this paper.

We further divide the clusters in redshift bins which coincide with the redshift limits of the slices used for the SZ map production.
This is to ensure that there is no bias to the pairwise kSZ signal on large scales.
Such a bias could arise because our way of slicing the simulation box does not preserve the large-scale correlations between adjacent slices.
The typical comoving thickness of a redshift slice is $\simeq 600~\Mpc$, whereas the largest scales at which we probe the pairwise kSZ signal are $300~\Mpc$.
Given that these scales only differ by a factor of two, the pairwise kSZ signal at the largest scales would be biased low because by construction we have removed the correlation between cluster pairs from different slices.
By only considering the signal in the appropriate redshift bins, we preclude such a bias. 

\begin{figure}
	\includegraphics[width=\columnwidth]{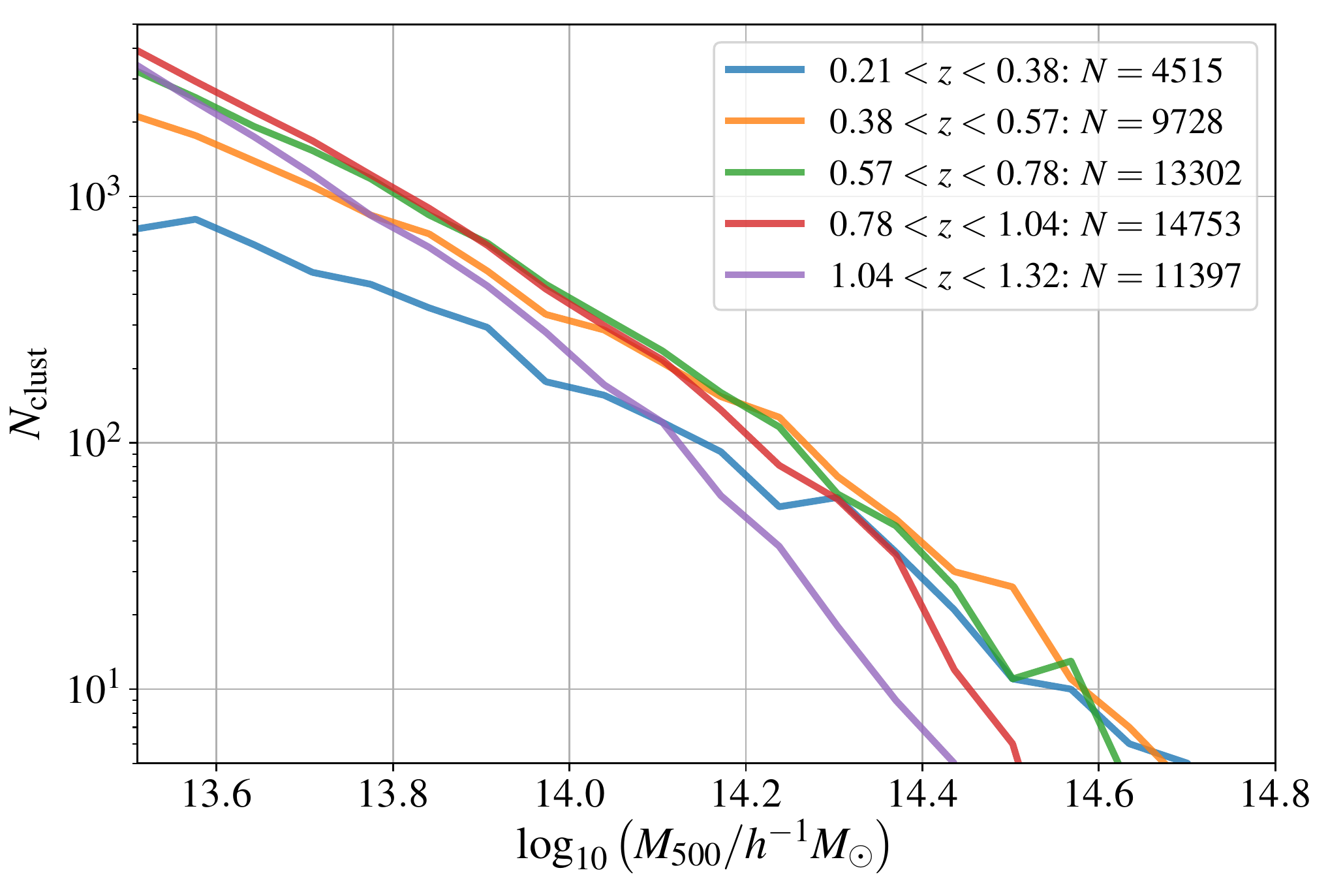}
	\caption{Mass distribution in the five redshift bins. Note that we plot the absolute number of clusters and not their comoving number density. The crossing of the distributions is a result of the combination of cluster growth and volume.
	}
	\label{fig:dNdM}
\end{figure}

We focus on the five redshift bins between $z = 0.21$ and \mbox{$z=1.32$}.
The excluded two lowest redshift bins ($z < 0.21$) have few clusters because of their small volume.
Similarly, the three highest bins ($z > 1.32$) would provide poor statistics because clusters are rare objects at these high redshifts.
The remaining redshift range $0.21 < z < 1.32$ is a good match to cluster samples expected from current and future surveys
and contains around 54,000 clusters with $M_{500c} > 3 \times 10^{13}~\msol$.
We show the mass distribution in these five bins in Fig.~\ref{fig:dNdM}.

\section{Analysis methods}
\label{sec:methods}

\subsection{Scaling relations for the optical depth}
The existence of a scaling relation between optical depth and tSZ signal was first pointed out by \cite{Diaferio2000}, who found that $ y \propto \tau^2$ was a good empirical fit to the results of their $N$-body simulations and semi-analytic modelling. 
\cite{Nagai2003} suggested reconstructing $\tau$ from the observed tSZ signal and mean electron temperature. However, this would require the density-weighted average $T_e$, which is not observable. 
Simulating one cluster at high resolution, they found that when using the emission-weighted temperature as measured from X-ray observations ($T_X$) instead, this leads to a significant bias ($\simeq 20\%$) on the recovered optical depth and therefore the peculiar velocity.
Using a sample of $\simeq 100$ massive clusters from a cosmological box, \cite{Diaferio2004} obtained similar conclusions, with estimated biases on the peculiar velocity ranging between $20$ and $50\%$.
These results were confirmed by \cite{Dolag2013} using more modern simulations. Furthermore, they found that this bias also propagates into the pairwise measurement: in their case the predicted $\tau$-values had to be reduced by $\simeq 30\%$ to match the pairwise kSZ signal as measured from the same simulation.
They also noted that, especially in massive systems, internal motions can contribute a non-negligible extra kSZ signal.

\citet[B16 hereafter]{Battaglia2016} used hydrodynamical simulations to calibrate a scaling relation between the optical depth and the Compton-$y$ parameter measured within the same aperture. 
This has the obvious advantage that no additional observational data are required: the tSZ signal and therefore $y$ can be measured from the same CMB survey as the kSZ.
B16 found a tight scaling relation with $\alpha \equiv \mathrm{d} \ln \tau / \mathrm{d} \ln y \simeq 0.5 $ for clusters with $M \gtrsim 10^{14} \Msol$ when averaged within fixed apertures between $1.3 \arcmin$ and $3.9\arcmin$.
\cite{deBernardis2016} compared the optical depth predicted from the B16 scaling relation to their result obtained with a pairwise kSZ template fit, finding acceptable agreement within the large uncertainties of their measurement.
\cite{Alonso2016} derived an analytic model for the $y$-$\tau$ relation. Based on the pressure profile of \cite{Arnaud2010} and the assumption of hydrostatic equilibrium they predict $\alpha \simeq 0.4$ when averaged within $R_{500}$. 
However, this approach does not take into account non-thermal pressure support, which can make up a non-negligible part of the total pressure in galaxy clusters (e.g.~\citealt{Battaglia2012,Nelson2014}).

On the other hand, \cite{Flender2017} calibrated a semi-analytic model for $\tau(M)$ on a compilation of high-resolution X-ray observations of in total $\simeq$~160 clusters.
Here the advantage is that one does not have to rely on hydrodynamical simulations with uncertain feedback models.
However, the calibrated model for the optical depth is only accurate if the clusters with high-resolution data are representative of the full population, both in terms of mass and redshift range and further properties (e.g.~cool-core vs. non-cool-core clusters).
Furthermore, the cluster mass is not a direct observable. When applying the $\tau(M)$ relation to a real cluster sample, any uncertainty in the mass-observable relation or in the estimate of hydrostatic mass bias will directly propagate into the prediction for $\tau$.

With these various options to predict the optical depth at hand, the question arises whether we can indeed compress the complex baryonic physics in galaxy clusters into a single number and use it to extract cosmology from future high-significance kSZ detections.
Here we therefore investigate whether a scaling relation for the optical depth can be used to recover an unbiased estimate of the mean pairwise velocity from future high-significance measurements of the pairwise kSZ signal.
A flow chart summarizing the analysis steps involved in this comparison is given in Fig.~\ref{fig:flow} below.
We focus on the $y$-$\tau$ relation, but we expect that similar conclusions can be drawn for alternative ways of predicting the optical depth.
When calibrating the $y$-$\tau$ relation, we extend the work of B16 to significantly lower masses and higher redshifts, and also point out some important subtleties in the calibration of this relation.
Throughout, the guiding principle is to produce a relation that can readily be applied to measurements of the pairwise kSZ signal from real data.

\subsection{Map filtering}
\label{subsec:mapfilt}

The observationally relevant quantities are the cylindrical Compton-$y$ parameter and optical depth integrated within an aperture $\theta_A$.
We measure these directly from the tSZ and kSZ maps produced from the simulation output as described in Sec.~\ref{subsec:maps}.
To facilitate the application of our scaling relation to real (beam-convolved) CMB data, we have convolved our kSZ and tSZ maps with a Gaussian beam with $\mathrm{FWHM} = 1.2\arcmin$, which is representative of high resolution CMB experiments like ACT and SPT. 
A more detailed discussion of the impact of the instrumental beam on the calibration of the scaling relation is given in Appendix~\ref{app:beam_impact}.

We then estimate the SZ signal of a cluster centred at position $\nhat_0$ as
\beq
T_\mathrm{filt}(\nhat_0) = \int d^2 \nhat \, T(\nhat) W(|\nhat - \nhat_0|) \, ,
\label{eq:realspacefilt}
\eeq
where $T(\nhat)$ is the tSZ or kSZ map, respectively.
The simplest possible choice is an average within the given aperture $\theta_A$, that is
\beq
W_\mathrm{avg}(\theta) = \frac{1}{\pi\theta_A^2} \times
\begin{cases}
	1, & \text{if $\theta < \theta_A$} \\
	0,  & \text{if $\theta > \theta_A$}
\end{cases}\, .
\label{eq:avgfilt}
\eeq
While this approach is sufficient for simulated tSZ- or kSZ-only maps, it is not well-suited for the application to real CMB data where the primary anisotropies on larger scales dominate the signal.
For the main analysis we therefore use an aperture photometry (AP) filter given by
\beq
W_\mathrm{AP}(\theta) =  \frac{1}{\pi\theta_A^2} \times
\begin{cases}
	1, & \text{if $\theta < \theta_A$} \\
	-1, & \text{if $\theta_A < \theta < \sqrt{2}\theta_A$}\\
	0,  & \text{if $\theta > \sqrt{2}\theta_A$} \, 
\end{cases}\, .
\label{eq:APfilt}
\eeq
The normalization constant ensures that this filter returns the average tSZ (or kSZ) signal within the aperture $\theta_A$, but corrected for contributions from larger scales.\footnote{The AP filter also efficiently removes scatter caused by the large-scale features in the kSZ map (see Fig.~\ref{fig:ksz_tsz_map}), which are caused by structures in the low-redshift slices of the simulation. 
}

The convolution of eq.~\ref{eq:realspacefilt} can be rewritten as a Fourier space filtering with $T_\mathrm{filt}(\bm{\ell}) = \tilde{W}(\ellta) T(\bm{\ell})$.
There are analytic expressions for the Fourier transforms of these filters given by
\begin{align}
\tilde{W}_\mathrm{avg}(\ellta) &= \frac{2}{\ellta}J_1(\ellta) \,  \\
\tilde{W}_\mathrm{AP}(\ellta) &= \frac{2}{\ellta} \left[2J_1(\ellta) - \sqrt{2} J_1(\sqrt{2}\ellta)\right] \, ,
\end{align}
respectively. 
Here $\ell$ is the modulus of the 2D wavenumber $\bm{\ell}$, and $J_n$ denotes a Bessel function of the first kind. 
As a consistency check for our pipeline, we have implemented and compared both the direct integration in pixel space and the Fourier space filtering.
While the Fourier space filtering is efficient in the case of a fixed aperture, it become computationally more expensive if an adaptive aperture is used.

The typical cluster size in our sample varies from $\theta_{500} \simeq 4 \arcmin$ at $M_{500c} \simeq 4\times 10^{14}~\hinv\Msol$ and $z \simeq 0.3$ to $\theta_{500} \simeq 0.7 \arcmin$ at $M_{500c} \simeq 4\times 10^{13}~\hinv \Msol$ and $z \simeq 1.3$.
A fixed angular aperture therefore probes different fractions of the cluster for a high-mass/low-$z$ object than for a low-mass/high-$z$ one.
Furthermore, the kSZ and tSZ signals have a different dependence on the angular distance from the cluster centre: the kSZ measures
$\int \dint l \, n_e$, whereas the tSZ measures $\int \dint l \, n_e T_e$, causing the latter to fall off more quickly towards the outskirts.

\begin{figure}
	\includegraphics[width=\columnwidth]{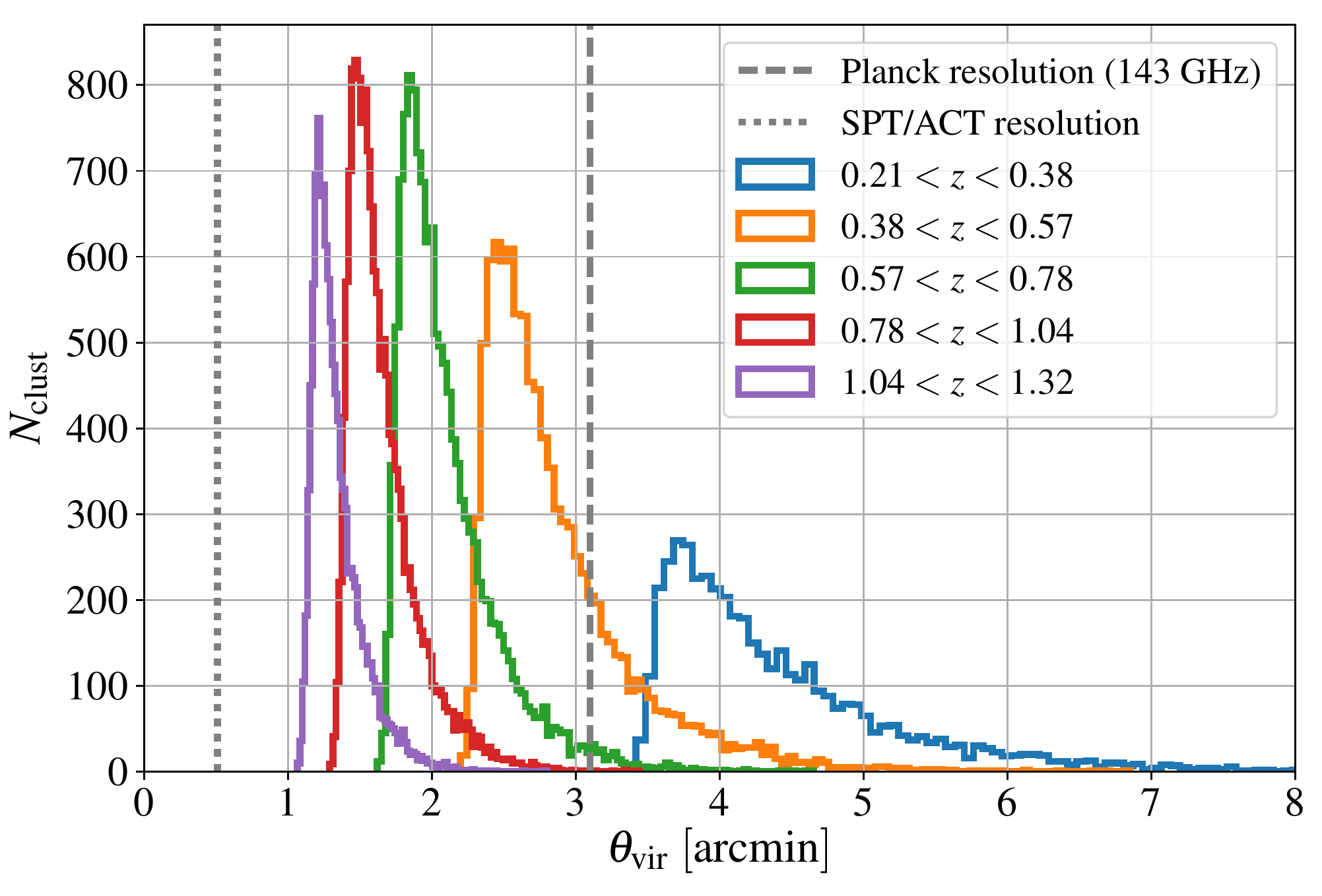}
	\caption{Angular size of the clusters: We show the distribution of $\theta_\mathrm{vir}$ for the five redshift bins considered in the analysis. We have also indicated the instrumental resolution of various experiments. Specifically, we show $\sigma_\mathrm{beam} = \mathrm{FWHM}/\sqrt{8\ln{2}}$ and have used $\mathrm{FWHM} = 7.3\arcmin$ for \Planck (143 GHz) and  $\mathrm{FWHM} = 1.2\arcmin$ for SPT/ACT.}
	\label{fig:thetavir_hist}
\end{figure}

For both of these reasons we use an adaptive aperture in our analysis.
In particular, we use the projected virial radius of the cluster, $\theta_\mathrm{vir} \equiv R_\mathrm{vir}/d_A$, where $d_A$ is the angular diameter distance.
The virial radius is given by
\beq
R_\mathrm{vir} = R_{500} \left( \frac{500 M_\mathrm{vir}}{\Delta_\mathrm{vir}(z) M_{500}} \right)^{1/3} \, ,
\eeq
with
$\Delta_\mathrm{vir}(z) = 18\pi^2 + 82\tilde{x} -39\tilde{x}^2$, 
$\tilde{x}= \Omega(z) -1$, and 
$\Omega(z) = \Omega_m(1+z)^3/E^2(z)$ \citep{BryanNorman98}.
This choice largely, but not entirely, avoids subtracting part of the signal from the outskirts in the aperture photometry filtering (see Sec.~\ref{subsec:meas_tau_y} and Appendix~\ref{app:tau_mgas} below for a more detailed discussion).
Furthermore, using the virial radius as an aperture minimizes the scatter in the relation between kSZ signal and halo bulk velocity \citep{Nagai2003}.
We show the distribution of angular radii for the five redshift bins in Fig.~\ref{fig:thetavir_hist}, and compare it to the angular resolution of \Planck and ground-based high-resolution CMB experiments (SPT, ACT). While most clusters at $z \gtrsim 0.4$ would be unresolved at the \Planck resolution, they are well resolved with SPT and ACT.

For real data $\theta_\mathrm{vir}$ will of course not readily be available. The average radius of a cluster at a certain redshift can, however, be predicted given a mass-observable relation (e.g.~\citealt{Saro2015,Simet2016,Melchior2016}).
We have verified that our results do not change significantly if we bin our clusters by their angular size and use the average radius within a bin as an aperture.
In other words, as long as the mass-observable relation is on average unbiased,
we do not require knowledge of the true angular size for every object.

To summarise, our AP filter returns a model-independent estimate for the average tSZ (kSZ) signal of any given cluster.
We note that a matched filter (e.g.~\citealt{Haehnelt1996}) could slightly increase the S/N, but it would require assuming a specific cluster profile. 
Furthermore, tSZ and kSZ have different angular dependencies (see eqs.~\ref{eq:ksz}-\ref{eq:tsz}), so a single matched filter would not be optimal for both signals relevant to our analysis.
For these reasons we use the more model-independent aperture photometry approach.

\subsection{$y$ and $\tau$ estimates}
\label{subsec:meas_tau_y}

\subsubsection{Measuring y}

\begin{figure}
	\includegraphics[width=\columnwidth]{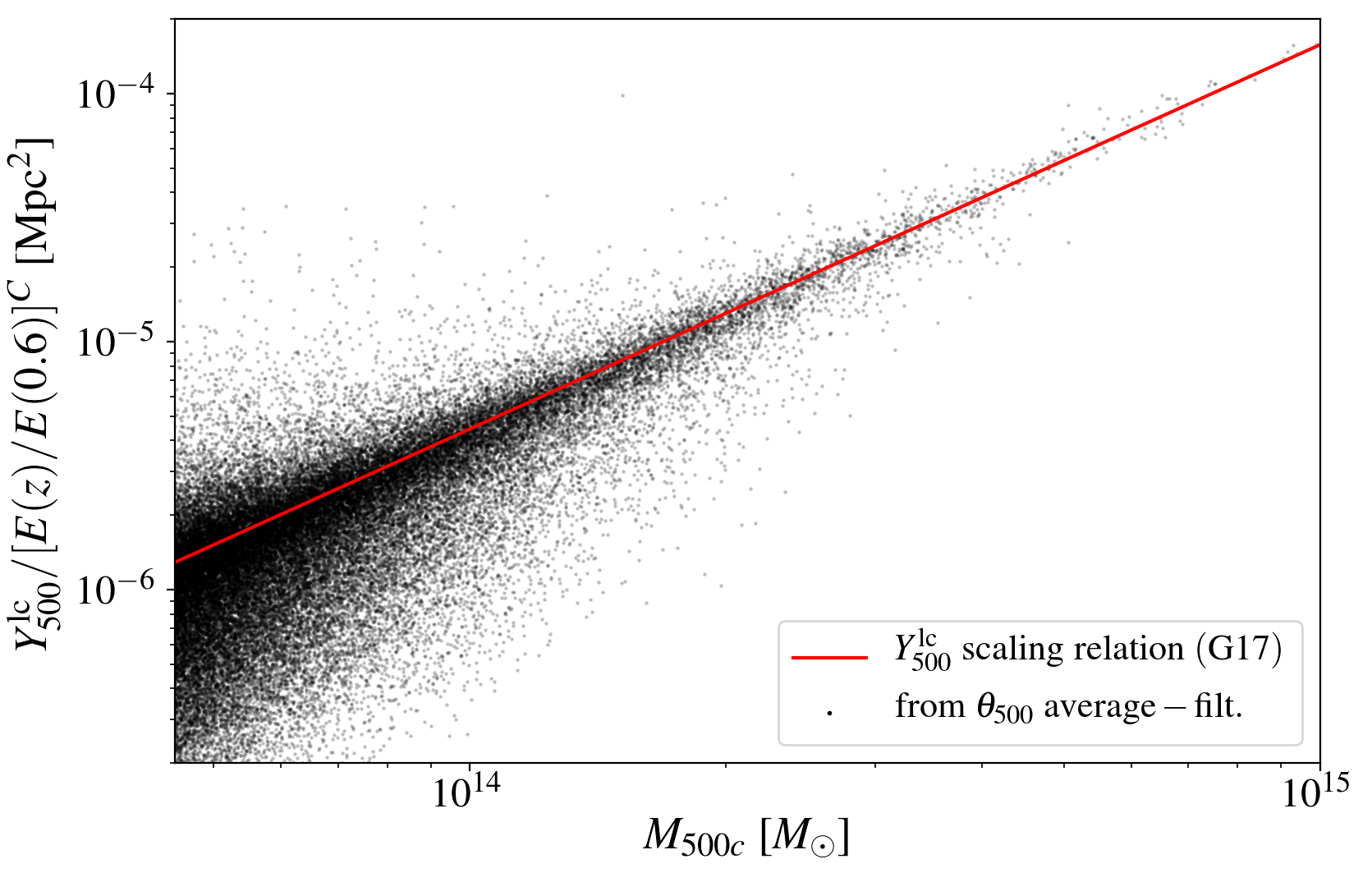}
	\caption{$Y$-$M$ relation: We show here the $Y^\mathrm{lc}_{500}$ values obtained by applying the average filter with $\theta_A = \theta_{500}$ to the $y$-map. 
	For comparison, we show in red the scaling relation fit by 
	\protect\cite{Gupta2016}
	using a smaller \textit{Magneticum} box and clusters with $M_{500} > 10^{14}~M_\odot$
	($Y^\mathrm{lc}_{500} \propto M_{500}^{1.55}$, redshift dependence $C=0.24$).
	The shallower slope of the $Y-M$ relation compared to the self-similar expectation ($Y \propto M^{5/3}$; \citealt{Kaiser1986}) is caused by the lightcone contribution to the observed signal, which affects low-mass haloes more strongly (see \citealt{Gupta2016} for details).
	}
	\label{fig:Ylc_test}
\end{figure}

The observed tSZ signal receives a lightcone contribution from other objects along the same line of sight.
This includes both correlated structure (the so-called 2-halo term) as well as chance associations along the line of sight.
By measuring the signal from the map (and not from the simulation particle data) we automatically take both of these contributions into account.
As a test for our pipeline, we first measure the integrated cylindrical tSZ signal within $\theta_{500}$ including the lightcone contribution, and compare to the results that \cite{Gupta2016} obtained from a smaller ($L = 896~\hinv \Mpc$) box of the \textit{Magneticum} suite and for clusters with $M_{500} > 10^{14}~M_\odot$.
For this purpose, we apply the average filter from eq.~\ref{eq:avgfilt} to the Compton-$y$ map produced from the simulations.
The integrated signal is then given by
\beq
Y^\mathrm{lc}_{500} = \langle y \rangle_{\theta_{500}} \times \pi (\theta_{500} d_A)^2 \,,
\eeq
where we denote the Compton-$y$ parameter averaged within an aperture $A$ as $\langle y \rangle_A$ (not to be confused with the average over the cluster sample denoted by $\overline{y}$).
We find excellent agreement between the two estimates, as shown in Fig.~\ref{fig:Ylc_test}.
We further note	that the \mbox{$Y$-$M$} scaling relation extends to masses well below $10^{14}~\Msol$ (down to where it was fitted), albeit with increasing scatter towards low masses.

We then proceed to measure the signal within the virial radius, $\langle y \rangle_\mathrm{vir}$, using the AP filter of eq.~\ref{eq:APfilt}.
We reiterate that the same AP procedure can be applied to observational data, as the AP also efficiently removes any fluctuations on scales larger than the aperture (such as the primary CMB).
Therefore we use $\langle y \rangle_\mathrm{vir}$ when calibrating the $y$-$\tau$ relation.

\subsubsection{Measuring $\tau$}

\begin{figure}
	\includegraphics[width=\columnwidth]{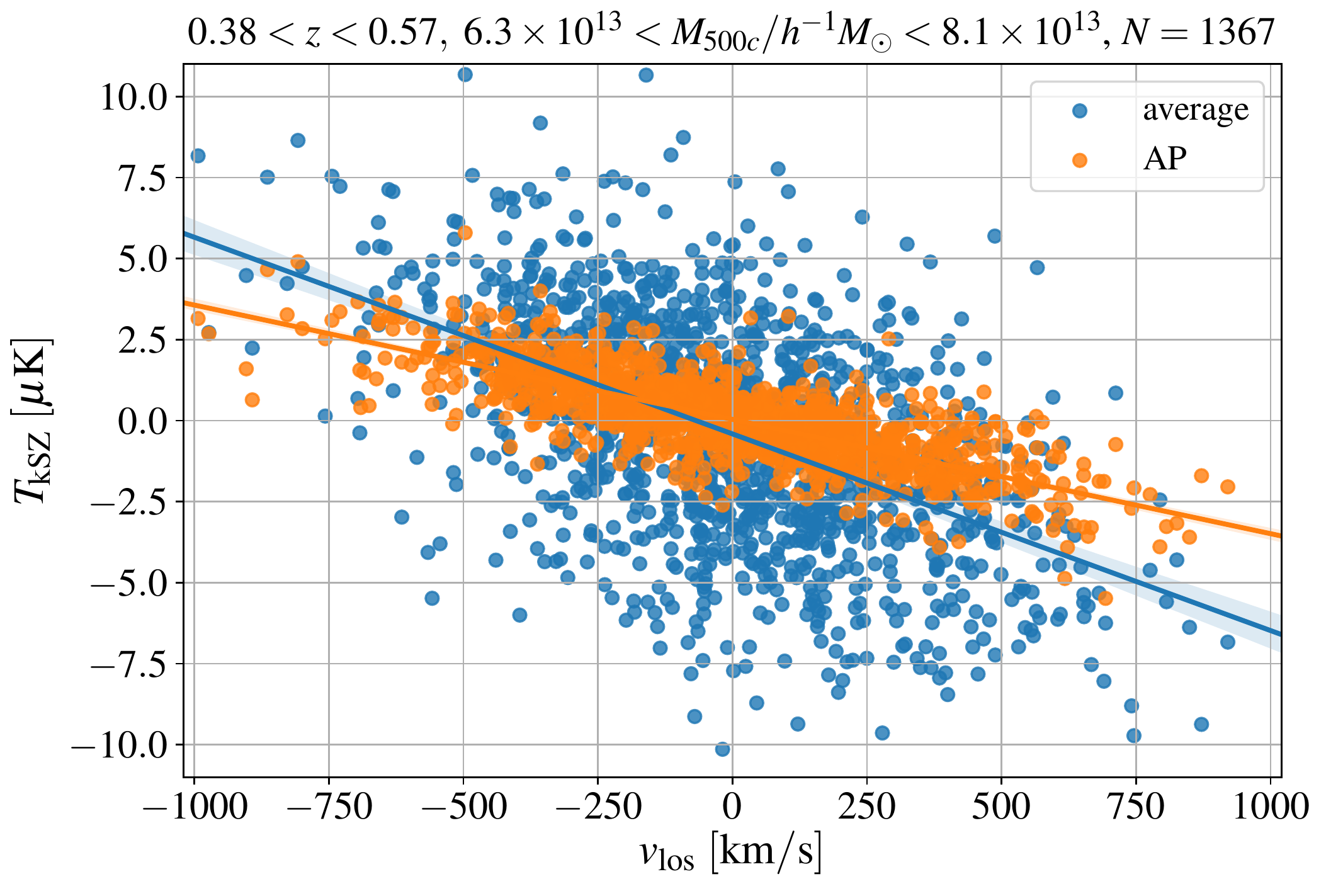}
	\caption{Determination of the optical depth: Here we show the tight relation between the AP-filtered $T_\mathrm{kSZ}$ and $\mathrm{v}_\mathrm{los}$ for one redshift slice and mass bin. The optical depth is determined from the slope of the relation using eq.~\ref{eq:ksz}. For illustration purposes we also show the same procedure using the average signal of every cluster (i.e.~without background subtraction), which leads to significantly larger scatter and a slightly steeper slope (see Appendix~\ref{app:tau_mgas}).
	}
	\label{fig:taufit}
\end{figure}

In B16 the optical depth of individual clusters was measured by projecting the electron density in the entire $165~\hinv\Mpc$ simulation box along one axis, and then averaging within a fixed aperture.
This approach works well enough for a small box, as the lightcone contribution is relatively small.
However, in our case it would lead to a significant bias of the measured optical depth from other structures along the same line of sight.

The kSZ signal at the position of individual clusters also receives a lightcone contribution, but when averaging over many clusters the latter averages to zero because of the velocity weighting in eq.~\ref{eq:ksz}.
Also, in the context of the pairwise kSZ measurement, we do not require the optical depth of any individual cluster, but only the mean $\overline{\tau}$ as a function of mass and redshift. 

We therefore measure $\overline{\tau}$ from the kSZ map as follows:
(1) For every cluster, we apply the adaptive AP filter with $\theta_A = \theta_\mathrm{vir}$ to the kSZ map to measure $\langle T_\mathrm{kSZ} \rangle_\mathrm{vir}$.
(2) In every redshift slice, we bin the clusters in 10 logarithmically spaced mass bins (here one could equally well bin by a mass proxy, or use a different number of bins).
(3) In every mass bin, we fit a linear relation between $ \langle T_\mathrm{kSZ} \rangle_\mathrm{vir} $ and the cluster line-of sight velocity $\mathrm{v}_\mathrm{los}$.
From the slope of this relation we determine  
$\overline{\langle \tau \rangle_\mathrm{vir}}$ and its associated uncertainty via eq.~\ref{eq:ksz}; an example for one redshift and mass bin is shown in Fig.~\ref{fig:taufit}.
This procedure, which is inspired by the velocity correlation method for kSZ detections, by construction yields the correct \textit{effective} optical depth of the cluster sample in the respective bin.
We note, however, that it is not a priori clear that this is also the correct quantity to use in eq.~\ref{eq:pksz} for the \textit{pairwise} kSZ measurement: as mentioned in Sec.~\ref{sec:intro} above, internal gas motions or correlations between velocity and optical depth could cause a bias.

In principle it is also possible to calculate the optical depth directly from the simulation particle data, as $\tau$ is directly related to the cylindrical gas mass (see Appendix \ref{app:tau_mgas}).
There are, however, a number of difficulties with this approach: 
(1) For the large simulation box and cluster sample considered here, it is computationally challenging to compute the cylindrical gas mass for every cluster.
(2) In addition, this calculation requires an arbitrary cut-off of the integration volume along the line of sight. Using too large a region would cause a bias from correlated structures along the same line of sight, whereas too small a region misses the outskirts of the cluster. 
(3) Including the effect of the beam convolution and aperture photometry on the measured optical depth (a non-negligible effect; see Appendix~\ref{app:beam_impact}), would require computing the projected gas density distribution of every cluster from the simulation data, which exacerbates the computational complexity. In our approach described above, on the other hand, inclusion of the beam is trivial. 

Nonetheless, we have computed the cylindrical gas mass directly from the simulation particle data for a representatively selected subset of clusters in a smaller simulation box (see Appendix~\ref{app:tau_mgas} for details).
We find that our AP estimate of the gas mass is on average $\simeq 25\%$ lower than the true $M_\mathrm{gas}^\mathrm{cyl}$ within the virial radius.
This difference arises because clusters are not perfect circles in projection and also do not have a sharp edge at $\theta_\mathrm{vir}$.
The AP estimate subtracts contributions from $\theta_\mathrm{vir} < \theta < \sqrt{2} \theta_\mathrm{vir}$ from the signal within $\theta_\mathrm{vir}$, thus reducing the measured gas mass.
As expected, the difference disappears when using the simple average-filter without background subtraction instead.
Therefore this comparison also serves as a consistency check for our measurement of the optical depth.

\subsection{$y$-$\tau$ scaling relation}
\label{subsec:scaling}

Following B16, we model the $y$-$\tau$ relation as
\beq
\ln \overline{\tau}  = \ln{\overline{\tau}_0} + \alpha \ln \frac{\overline{y} }{\overline{y}_0} \,,
\label{eq:scalingrel}
\eeq
where we have set the normalization constant to $\overline{y}_0 = 10^{-6}$. This is a typical value for the aperture-averaged Compton-$y$ parameter for the clusters in our sample, so that $\overline{\tau}_0$ is also representative for the typical aperture-averaged optical depth.\footnote{We will omit the $\langle . \rangle_\mathrm{vir}$ unless there is a potential ambiguity.}

\begin{figure}
	\includegraphics[width=\columnwidth]{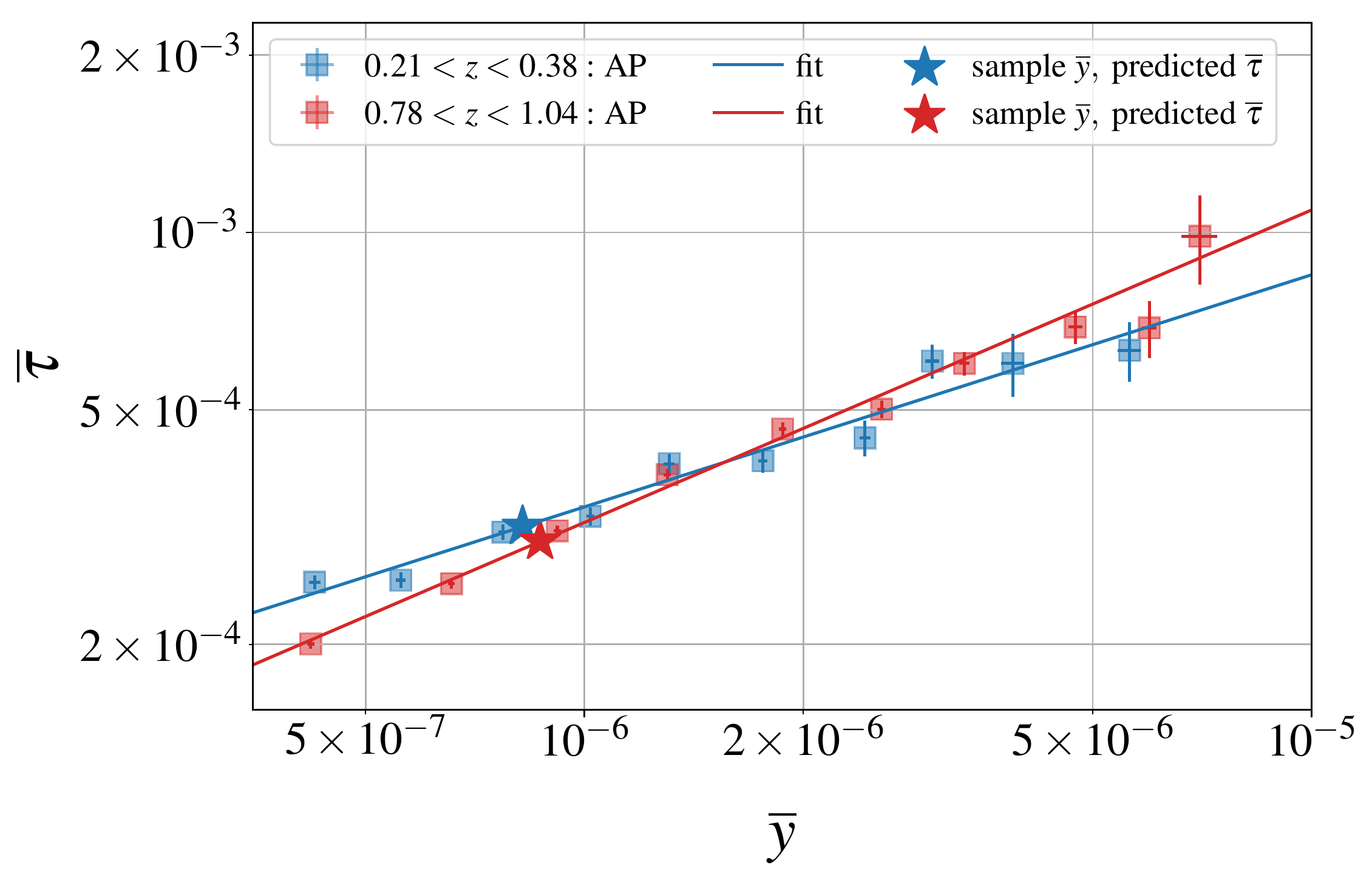}
	\caption{$\overline{y}$-$\overline{\tau}$ scaling relation: We show here the scaling relation for two redshift slices as an example. The square points
	with error bars denote the individual mass bins of the respective slice; the solid line is the scaling relation fit to them. The position of the asterisk
	shows the average $\bar{y}$ across all clusters in the slice, and the corresponding $\overline{\tau}$ predicted from the scaling relation.
	}
	\label{fig:y_tau_scaling}
\end{figure}

Using the binned $\overline{\tau} $ estimates and the $\overline{y}$ computed in the same bins, we then fit for $\alpha$ and  $\ln \overline{\tau}_0$ using weighted orthogonal distance  regression\footnote{\url{https://docs.scipy.org/doc/scipy/reference/odr.html}}.
We repeat this procedure in every redshift slice, and show an example for two slices in Fig.~\ref{fig:y_tau_scaling}. 
The resulting fit parameters are reported in Table~\ref{tab:fitparams}.
We have also included the off-diagonal element of the correlation matrix $\mathrm{Corr}(\alpha,\ln \overline{\tau}_0) = C_{\alpha,\ln \overline{\tau}_0}/(\sigma_\alpha \sigma_{\ln \overline{\tau}_0})$, 
where $C_{\alpha,\ln \overline{\tau}_0}$ is the parameter covariance obtained from the fit.

We note that while there is little change in the normalization $\ln \overline{\tau}_0$, the slope evolves from $\alpha \simeq 0.4$ at low $z$ to $\alpha \simeq 0.6$ at high $z$.
It is instructive to compare these values to the expectation assuming self-similarity. 
In this case one would expect $\overline{y} \propto Y/\theta_\mathrm{vir}^2 \propto M^{5/3}/M^{2/3} = M$ for the average Compton-$y$ parameter, and similarly  $\overline{\tau} \propto \mathcal{T}/\theta_\mathrm{vir}^2 \propto M/M^{2/3} = M^{1/3}$ (with $Y$ and $\mathcal{T}$ being the Compton-$y$ parameter and optical depth integrated over the aperture).
The self-similar expectation would therefore be $\overline{\tau} \propto \overline{y}^{1/3}$.
At low redshift we measure an exponent that is relatively close to this value, while at higher redshifts we find a significantly steeper relation.

We have also computed the power-law exponent of the \mbox{$Y-M$} and $\mathcal{T}-M$ relations;
here we use the integrated quantities to avoid an impact of the changing angular diameter distance on the scaling relations.
We obtain $Y \propto M^{1.72 \pm 0.01}$ and $\mathcal{T} \propto M^{1.09 \pm 0.03}$ in the lowest redshift bin, which is again close to the self-similar expectation ($Y \propto M^{5/3}$, $\mathcal{T} \propto M$).
Both scaling relations steepen towards higher redshift, but the trend is stronger in the $\mathcal{T}-M$ relation, leading to the observed steepening in the $\overline{\tau}$-$\overline{y}$ relation.
A possible physical explanation for these trends might be the impact of AGN feedback breaking self-similarity at higher redshifts.
\footnote{
We note, however, that the subtraction of signal from the cluster outskirts during the AP filtering (see Appendix~\ref{app:tau_mgas}) could also have a small impact on the exponent of these relations.
}

\begin{table*}
	\begin{tabular}{ccccccccc}
\toprule
Parameter &      0.21 < z < 0.38 &      0.38 < z < 0.57 &      0.57 < z < 0.78 &      0.78 < z < 1.04 &      1.04 < z < 1.32 \\
\midrule
$\alpha$                                     &      $0.39 \pm 0.03$ &      $0.40 \pm 0.01$ &      $0.46 \pm 0.01$ &      $0.53 \pm 0.02$ &      $0.59 \pm 0.02$ \\
$\mathrm{ln}~\overline{\tau}_0$              &     $-7.98 \pm 0.02$ &     $-7.94 \pm 0.01$ &     $-7.98 \pm 0.01$ &     $-8.04 \pm 0.01$ &     $-8.13 \pm 0.01$ \\
Corr($\alpha,\mathrm{ln}~\overline{\tau}_0)$ &               $0.22$ &               $0.10$ &              $-0.01$ &               $0.00$ &               $0.07$ \\
$\chi^2/\mathrm{dof}$                        &               $2.06$ &               $0.42$ &               $0.66$ &               $1.74$ &               $0.98$ \\
$\overline{y}$                               &  $8.22\times10^{-7}$ &  $8.94\times10^{-7}$ &  $9.22\times10^{-7}$ &  $8.69\times10^{-7}$ &  $7.67\times10^{-7}$ \\
predicted $\overline{\tau}$                  &  $3.17\times10^{-4}$ &  $3.42\times10^{-4}$ &  $3.30\times10^{-4}$ &  $2.99\times10^{-4}$ &  $2.52\times10^{-4}$ \\
$\sigma_{\overline{\tau}}/\overline{\tau}$   &                  2\% &                  1\% &                  1\% &                  1\% &                  2\% \\
\bottomrule
\end{tabular}

	\caption{Parameters of the $\overline{\tau}$-$\overline{y}$ scaling relation: We show here the results from the fit together with the mean $\overline{y}$ and predicted $\overline{\tau}$ in every redshift bin.
		While the first and fourth redshift slice show a somewhat high $\chi^2$ when fitting the scaling relation, the overall $\chi^2$ is acceptable ($\chi^2_\mathrm{all-z} = 47$ for 40 degrees of freedom, probability to exceed of $21\%$). 
	}
	\label{tab:fitparams}
\end{table*}

Provided a measurement of the average tSZ signal of a cluster sample, eq.~\ref{eq:scalingrel} can now be used to predict the optical depth within the same aperture.
We compute the predictions for $\overline{\tau}$ separately for all redshift bins in our sample and also report them in Table~\ref{tab:fitparams}.
We will later use these values to convert our pairwise kSZ measurement into an estimate for the mean pairwise velocity.
Propagating the statistical uncertainty from the fit and the $\overline{y}$ measurement, the relative error on the predicted $\overline{\tau}$ is given by
\beq
\frac{\sigma_{\overline{\tau}}}{\overline{\tau}} = 
\left[ 
\sigma^2_{\ln \overline{\tau}_0}
+ 2 \ln \left(\frac{\overline{y}}{\overline{y}_0}\right) C_{\alpha,\ln \overline{\tau}_0}   
+ \left(\ln\frac{\overline{y}}{\overline{y}_0}\right)^2 \sigma_\alpha^2 
+ \alpha^2 \left(\frac{\sigma_{\overline{y}}}{\overline{y}}\right)^2
\right]^{1/2} \,.
\label{eq:tau_error}
\eeq
In the case of a perfectly clean $y$-map the fractional uncertainty is only 1 to $2\%$ (see Table~\ref{tab:fitparams}), well below the systematic uncertainty introduced by the sub-grid feedback models used in the simulation (see Sec.~\ref{subsec:systematics} below for a further discussion).
Applied to real data the uncertainty of the $\overline{y}$-measurement would propagate into the prediction for $\overline{\tau}$.

\subsection{Pairwise estimator}
\label{subsec:pairwise}

We use the estimator for the mean pairwise velocity derived by \cite{Ferreira1998}, 
\beq
\hat{\mathcal{v}}_{12}(r) = \frac{\sum_{i<j,r} (\hat{\mathbf{r}}_i \cdot \mathbf{v}_i -\hat{\mathbf{r}}_j \cdot \mathbf{v}_j ) \, c_{ij}} {\sum_{i<j,r} c_{ij}^2} \, .
\label{eq:ferreira}
\eeq
Here the weights are given by
\beq
c_{ij} = \hat{\mb{r}}_{ij} \cdot \frac{\hat{\mb{r}}_i + \hat{\mb{r}}_j}{2} \, \quad \text{with} \quad \mb{r}_{ij} \equiv \mb{r}_i - \mb{r}_j \, , \quad |\mb{r}_{ij}| = r \,.
\eeq

As the kSZ effect provides a proxy for the line-of-sight velocity (see~eq.~\ref{eq:ksz}), the same estimator can be applied to kSZ temperature maps \citep{Hand2012}:
\beq
\hat{T}_{\mathrm{pkSZ}}(r) = - \frac{\sum_{i<j, r} \left[ T(\nhat_i)-T(\nhat_j) \right ] c_{ij}}{\sum_{i<j,r} c_{ij}^2} \,.
\label{eq:pkszest}
\eeq
Here we use the same AP-filtered kSZ temperatures as before.

Comparing the estimators for $\mathcal{v}_{12}$ and $T_\mathrm{pkSZ}$ (eqs.~\ref{eq:ferreira} and~\ref{eq:pkszest}), it becomes apparent why eq.~\ref{eq:pksz} is an approximation whose accuracy needs to be tested.
Schematically, we have 
\begin{align}
\begin{split}
- \left\langle T_1 - T_2 \right \rangle_c &\simeq 
\frac{T_\mathrm{CMB}}{c} \left\langle \tau_1 v^\mathrm{los}_1 - \tau_2 v^\mathrm{los}_2 \right \rangle_c \\
&\simeq \frac{T_\mathrm{CMB}}{c} \overline{\tau}\left\langle v^\mathrm{los}_1- v^\mathrm{los}_2 \right\rangle_c 
= \overline{\tau} \frac{T_\mathrm{CMB}}{c} \mathcal{v}_{12}
\, ,
\end{split}
\end{align}
where we have used the shorthand $\langle ... \rangle_c$ for the estimator of eqs.~\ref{eq:ferreira} and \ref{eq:pkszest}.	
The first approximate equality assumes that internal cluster motions do not introduce a bias to the signal.
The second step assumes that there is no strong correlation between optical depth and velocity.

We follow previous studies (e.g.~\citealt{Hand2012,Planck_KSZ}) and 
subtract the mean temperature as a function of redshift from the kSZ temperature data, i.e.~
\beq
\label{eq:T_zevolcorr}
T_i^\mathrm{corr} = T_i - \overline{T}(z_i) = T_i - \frac{ \sum_j T_j \, G(z_i,z_j,\Sigma_z)}{\sum_j G(z_i,z_j,\Sigma_z)} \, ,
\eeq
where $G(z_i,z_j,\Sigma_z) = \exp\left[-0.5(z_i-z_j)^2/\Sigma_z^2 \right]$. 
We set $\Sigma_z = 0.01$ to obtain a smoothly varying mean temperature.
Furthermore, we apply the same correction to the line-of-sight velocities in the halo catalogue.
This is necessary because otherwise large-scale velocity modes within a finite slice through the simulation can cause a bias to the signal on large scales.
We then apply this estimator in 20 separation bins linearly spaced between 0 and 300~Mpc.

We further estimate the uncertainties on the pairwise kSZ measurement by bootstrap resampling from the cluster catalogue.
Having created $N_\mathrm{BS}$ bootstrap realisations by drawing with replacement from the catalogue,
the covariance matrix is given by
\beq
\hat{C}_{ij}^\mathrm{BS} = (N_\mathrm{BS}-1)^{-1} \, \sum_{\alpha = 1}^{N_\mathrm{BS}} (\hat{T}_i^\alpha - \bar{T}_i) \, (\hat{T}_j^\alpha - \bar{T}_j) \, .
\label{eq:C_bootstrap}
\eeq
Here $ \hat{T}_i^\alpha $ is the pairwise signal in separation bin $i$ and bootstrap realisation $\alpha$, and $\bar T_i$ is its mean over all realisations.
We use $N_\mathrm{BS} = 2000$ resamples; our covariance is stable against a further increase.\footnote{
We have also computed the covariance from jackknife resamples and find good agreement between the two estimates. For further discussion of covariance estimates for the pairwise kSZ measurement see \cite{Soergel2016} and \cite{deBernardis2016}.
}
When using the inverse covariance matrix $C^{-1}$, we include the correction factor by \cite{Hartlap2006} to account for the fact that $(C^\mathrm{BS})^{-1}$ is a biased estimator of $C^{-1}$.

\begin{figure}
	\includegraphics[width=\columnwidth]{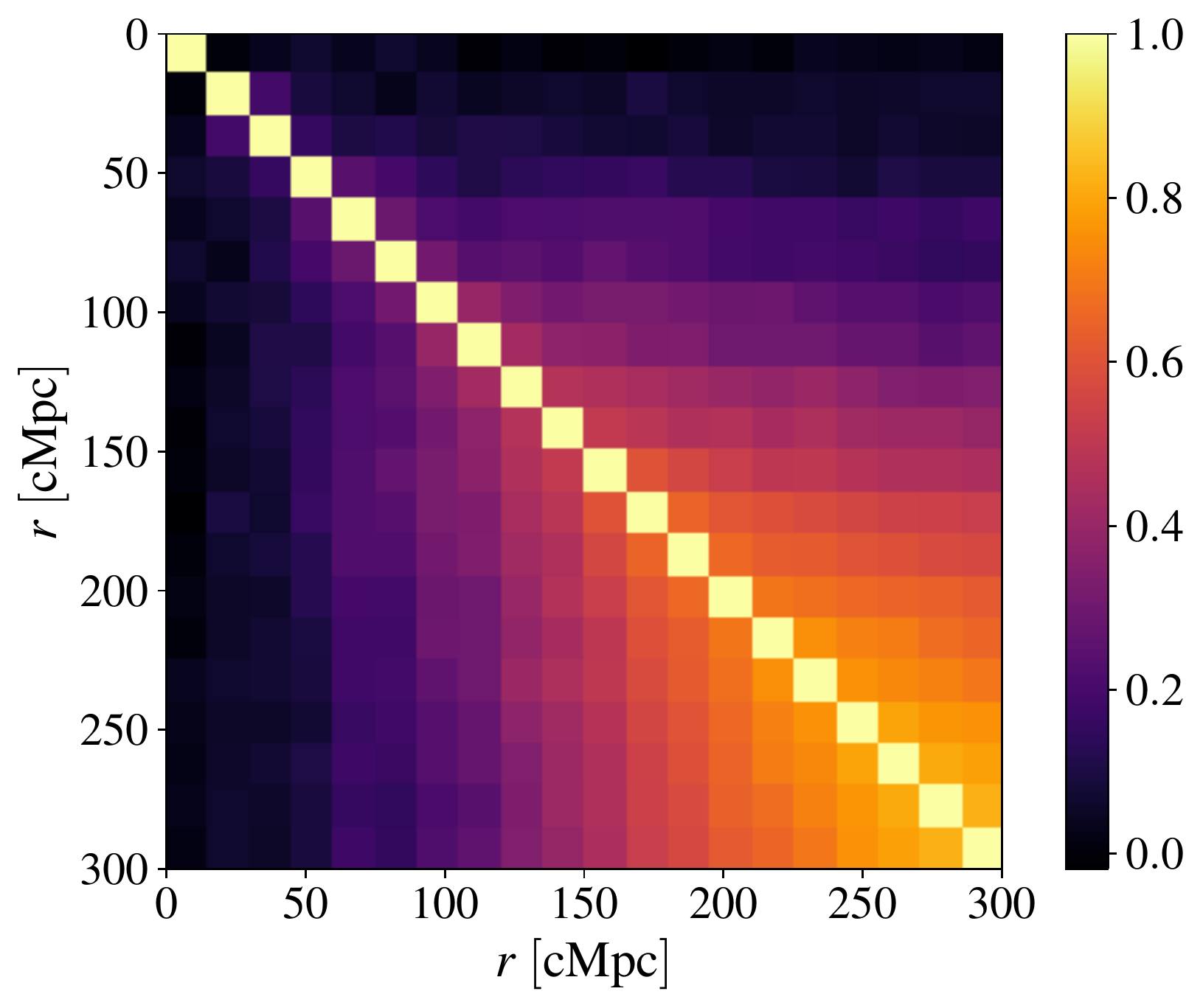}
	\caption{Correlation matrix of the pairwise kSZ measurement in the redshift bin $0.57 < z < 0.78$
		as estimated from bootstrap resampling from the catalogue (eq.~\ref{eq:C_bootstrap}).}
	\label{fig:corrmat}
\end{figure}

For visualisation purposes, we further define the correlation matrix $R_{ij} = C_{ij}/\sqrt{C_{ii} C_{jj}}$ (no sum over repeated indices) and show an example for the central redshift slice in Fig.~\ref{fig:corrmat}.
For separations $r \gtrsim 100~\Mpc$ there are significant correlations between adjacent separation bins.

\subsection{Pairwise velocity}
\label{subsec:v12}

We convert the pairwise kSZ signal into an estimate of the mean pairwise velocity, 
$\mathcal{v}_{12}^\mathrm{kSZ}$,
via eq.~\ref{eq:pksz}. Its covariance matrix is given by
\beq
C^{\mathcal{v}_{12}}_{ij} = C^\mathrm{BS}_{ij} \left(\frac{c}{\overline{\tau}T_\mathrm{CMB}}\right)^2 + \left(\frac{\sigma_{\overline{\tau}}}{\overline{\tau}}\right)^2 \mathcal{v}_{12,i}^\mathrm{kSZ} \mathcal{v}_{12,j}^\mathrm{kSZ} \,
\eeq
where the second term propagates the uncertainty in the estimate of $\overline{\tau}$.
Because we are comparing within the same simulation, we do not include the systematic uncertainty due to the sub-grid feedback models (see Sec.~\ref{subsec:systematics} below) into $\sigma_{\overline{\tau}}/\overline{\tau}$  here. It should, however, be included when using the scaling relation on real data.

We finally assess how well we recover the true pairwise velocity as measured directly from the halo catalogue.
Given a measured data vector $\hat{\mb{s}}$ with inverse covariance matrix $C^{-1}_s$ and a theoretical model $\mb{s}_\mathrm{th}$, we define
\beq
\chi^2_a \equiv (\hat{\mb{s}} - a \mb{s}_\mathrm{th})^T C_s^{-1} (\hat{\mb{s}} - a \mb{s}_\mathrm{th}) \, .
\eeq
The best-fitting value of $a$, its uncertainty, and the statistical significance are then given by
\beq
a = \frac{\hat{\mb{s}}^T C_s^{-1} \mb{s}_\mathrm{th}}{\mb{s}_\mathrm{th}^T C_s^{-1} \mb{s}_\mathrm{th}} \,, \qquad 
\sigma_a^2 = \frac{1}{\mb{s}_\mathrm{th}^T C_s^{-1} \mb{s}_\mathrm{th}} \, ,
\qquad \frac{S}{N} = \frac{a}{\sigma_a} \, ,
\label{eq:templatefit}
\eeq
respectively.
We have defined the amplitude parameter such that $a = 1$ indicates perfect agreement between model and measurement.
Using $\mathcal{v}_{12}$ measured directly from the halo catalogue as the theoretical template, we 
quantify the agreement between recovered and true mean pairwise velocity via eq.~\ref{eq:templatefit}. \footnote{
If instead we want to be agnostic about the expected shape of the signal, we define $ \chi^2_\mathrm{null} \equiv \hat{\mb{s}}^T C_s^{-1} \hat{\mb{s}}$. We then convert the associated probability to exceed (PTE) of the $\chi^2$-distribution with 20 degrees of freedom to a significance in standard deviations via the standard normal distribution. In the limit of high $S/N$ (as it is the case for our kSZ only-measurement) the two approaches yield very similar answers.
}

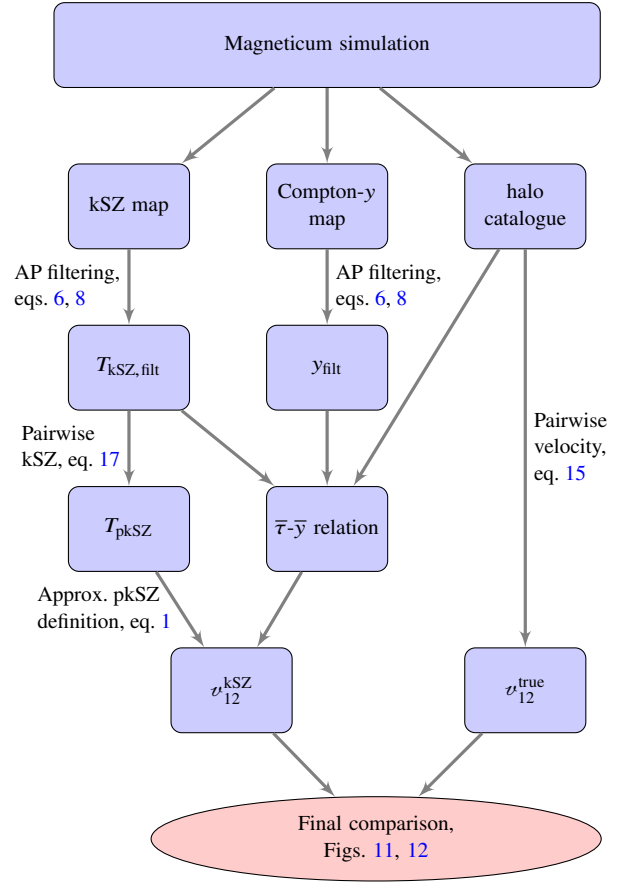
\begin{figure}
	\tikzstyle{decision} = [diamond, draw, fill=blue!20,
	text width=4.5em, text badly centered, node distance=2.5cm, inner sep=0pt]
	\tikzstyle{block} = [rectangle, draw, fill=blue!20,
	text width=5em, text centered, rounded corners, minimum height=4em]
	\tikzstyle{line} = [draw, very thick, color=black!50, -latex']
	\tikzstyle{dashedline} = [draw, dashed, very thick, color=black!50, -latex']
	\tikzstyle{dotline} = [draw, dotted, very thick, color=black!50, -latex']
	\tikzstyle{cloud} = [draw, ellipse,fill=red!20, node distance=2.5cm,
	minimum height=2em]
	\usetikzlibrary{positioning}
	\begin{center}
		\begin{tikzpicture}[scale=2, node distance = 1cm, auto]
		\node [block, text width=7cm, align=center] (init) {Magneticum simulation};
		\node [block, below = of init] (tSZmap) {Compton-$y$ map};
		\node [block, left = of tSZmap] (kSZmap) {kSZ map};
		\node [block, below = of kSZmap] (TkSZfilt) {$T_{\mathrm{kSZ,filt}}$};
		\node [block, below = of TkSZfilt] (TpkSZ) {$T_{\mathrm{pkSZ}}$};
		\node [block, below = of tSZmap] (TtSZfilt) {$y_{\mathrm{filt}}$};
		\node [block, below = of TtSZfilt] (taubar) {$\overline{\tau}$-$\overline{y}$ relation};
		\node [block, below left = 1cm and -0.35cm of taubar] (v12) {$\mathcal{v}_{12}^\mathrm{kSZ}$};
		\node [block, right = of tSZmap] (halocat) {halo catalogue};
		\node [block, below right = 1cm and 1cm of taubar] (v12exact) {$\mathcal{v}_{12}^\mathrm{true}$};
		\node [cloud, below right = 1cm and -1cm of v12, text width=4cm, align=center] (final) {Final comparison,\\Figs.~\ref{fig:result_zbins},~\ref{fig:result_zbins_ratio}};
		\path [line] (init) -- node [left, color=black] {} (kSZmap);
		\path [line] (init) -- node [left, color=black] {} (tSZmap);       
		\path [line] (init) -- node [left, color=black] {} (halocat);
		\path [line] (kSZmap) -- node [left, color=black, align=left] {AP filtering,\\eqs.~\ref{eq:realspacefilt}, \ref{eq:APfilt}}(TkSZfilt);
		\path [line] (tSZmap) -- node [, color=black, align=left] {AP filtering,\\eqs.~\ref{eq:realspacefilt}, \ref{eq:APfilt}}(TtSZfilt);
		\path [line] (TkSZfilt) -- node [left, color=black, align=left] {Pairwise\\ kSZ, eq.~\ref{eq:pkszest}}(TpkSZ);
		\path [line] (TtSZfilt) -- node [right, color=black, align=left] {}(taubar);
		\path [line] (TkSZfilt) -- node [right, color=black, align=left] {}(taubar); 
		\path [line] (halocat) -- node [right, color=black, align=left] {}(taubar); 
		\path [line] (TpkSZ) -- node [, left, color=black, align=left] {Approx. pkSZ\\ definition, eq.~\ref{eq:pksz}}(v12);
		\path [line] (taubar) -- node [, color=black] {}(v12);
		\path [line] (halocat) -- node [, color=black, align=left] {Pairwise\\velocity,\\eq.~\ref{eq:ferreira}}(v12exact);
		\path [line] (v12) -- node [left, color=black, align=left] {}(final);
		\path [line] (v12exact) -- node [left, color=black] {}(final);
		\end{tikzpicture}
		\caption{Flowchart summarising the calibration and validation procedure used in this work.
		}
		\label{fig:flow}
	\end{center}
\end{figure}

\section{Results and discussion}
\label{sec:results}

\subsection{Pairwise kSZ vs. pairwise velocity}
\begin{figure}
	\includegraphics[width=\columnwidth]{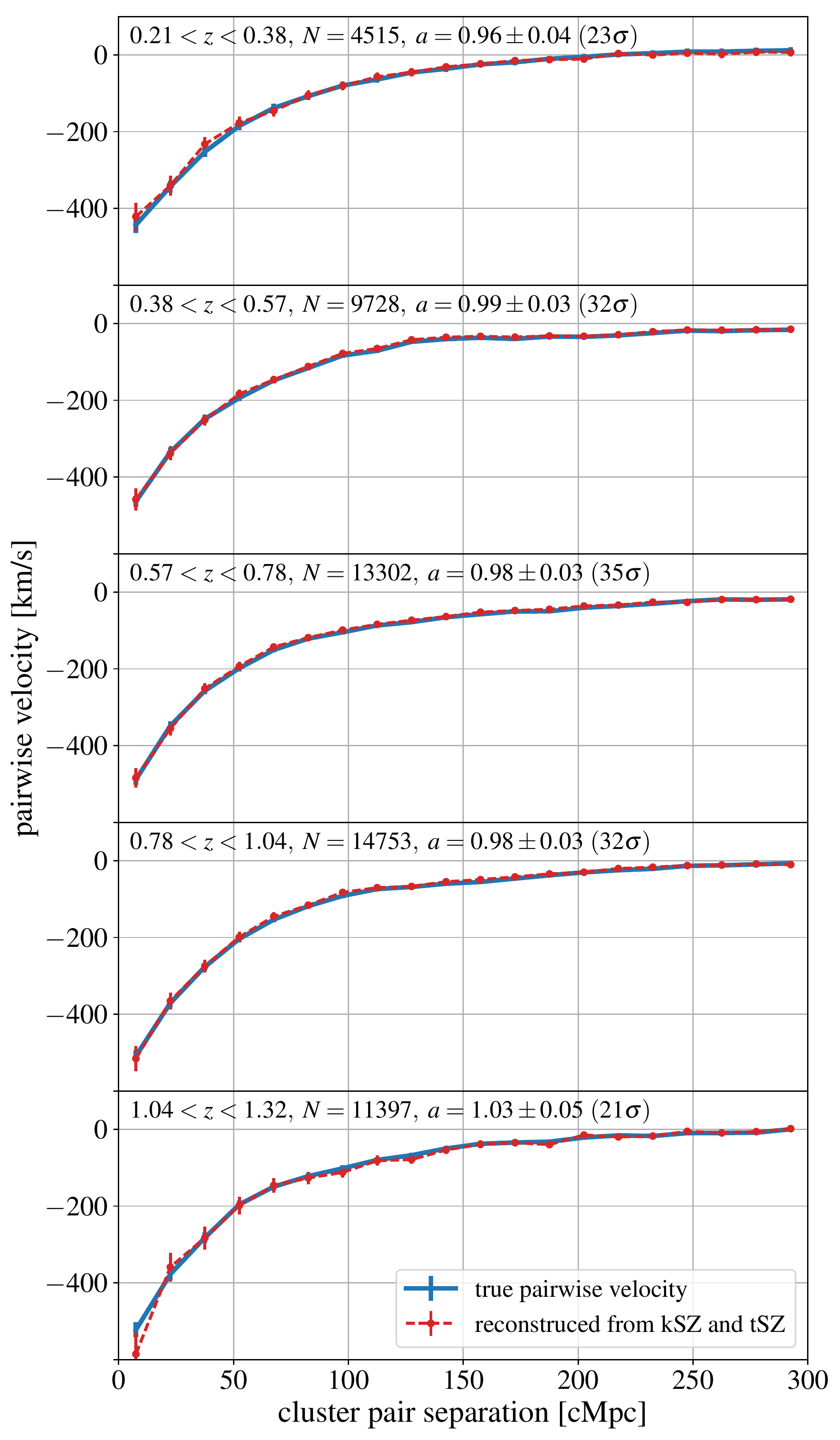}
	\caption{Mean pairwise velocity: We show here
		$\mathcal{v}_{12}^\mathrm{kSZ}$ (obtained from the pairwise kSZ signal and the $y$-$\tau$ scaling relation)
		compared to the true values computed directly from the halo catalogue. The error bars on 
		$\mathcal{v}_{12}^\mathrm{kSZ}$
		include the statistical errors of the scaling relation, but not the systematic uncertainty due to sub-grid feedback models (see Secs.~\ref{subsec:scaling} and~\ref{subsec:systematics}).
		The individual panels show the signal in the different redshift bins. The number of clusters in every bin is given in the respective panel; 
		we have also included the best fitting amplitude and statistical significance (see Sec.~\ref{subsec:v12}).
	}
	\label{fig:result_zbins}
\end{figure}

\begin{figure}
	\includegraphics[width=\columnwidth]{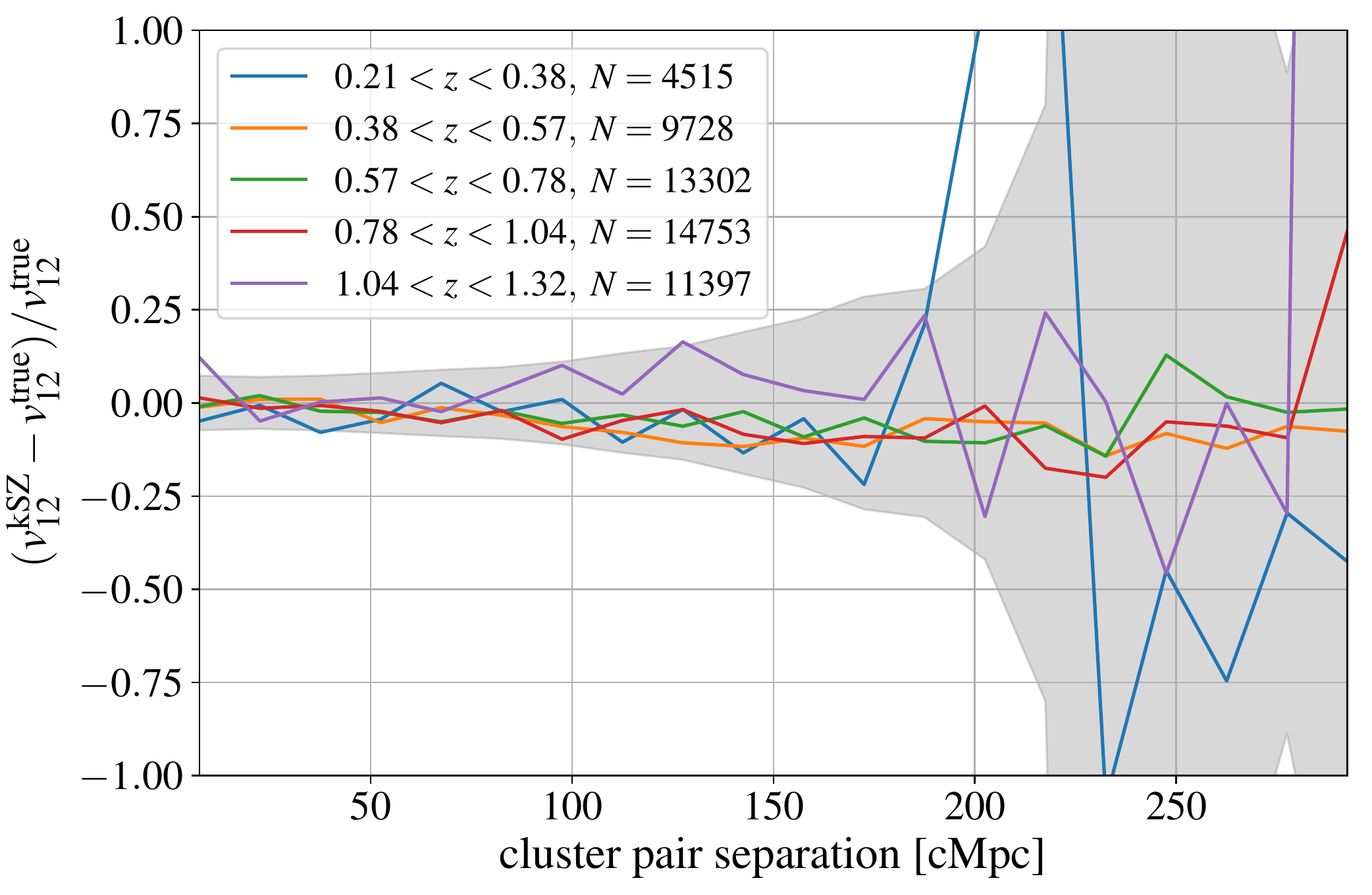}
	\caption{Fractional difference between the mean pairwise velocity recovered from the pairwise kSZ signal and the true values. For reference, the grey shaded band shows the fractional statistical uncertainty on $\mathcal{v}_{12}^\mathrm{kSZ}$ as a function of scale (averaged over all five redshift bins). The increase in fractional error towards large separations is a result of $\mathcal{v}_{12}$ going towards zero at large $r$.
	}
	\label{fig:result_zbins_ratio}
\end{figure}

The underlying assumption of using the pairwise kSZ signal as a probe of cosmology is the simple multiplicative relation given by eq.~\ref{eq:pksz}.
Having established the $\tau$-$y$ scaling relation and the pairwise kSZ estimate, we can now test whether this assumption holds at the high precision required by future surveys.

We compare the estimate of $\mathcal{v}_{12}$ from the pairwise kSZ signal (eq.~\ref{eq:pkszest}) and the $\tau$-$y$ scaling relation to the result obtained directly from the halo catalogue (eq.~\ref{eq:ferreira}).
An overview of the various steps involved in this comparison is given in Fig.~\ref{fig:flow}.
Its results are shown in Fig.~\ref{fig:result_zbins}, 
where we have also included the amplitude fits and statistical significance as computed from eq.~\ref{eq:templatefit}.
We find that the pairwise kSZ signal combined with the $\tau$-$y$ scaling relation
yields an accurate estimate of the true mean pairwise velocity.

The amplitude of the recovered signal is consistent with unity within the statistical uncertainties in all redshift bins. The significances range between $21\sigma$ in the highest redshift bin ($1.04 < z < 1.32$) and $35\sigma$ in the central bin ($0.57 < z < 0.78$). 
As we have assumed a perfect kSZ map and a catalogue containing every cluster with $M_{500c} > 3 \cdot 10^{13}~\msol$, these numbers are higher than more elaborate forecasts for future experiments (e.g.~\citealt{Flender2015}).
They do, however, highlight that there is a very significant pairwise kSZ signal, especially when pushing down towards lower mass limits.

We further show the ratio between $\mathcal{v}_{12}$ recovered from the pairwise kSZ and the true values in Fig~\ref{fig:result_zbins_ratio}.
In all redshift bins, and at all scales  $r \lesssim 150~\Mpc$, the differences between the recovered and true $\mathcal{v}_{12}$ are well below $10\%$.
Finally, the difference between true and recovered signal is also well below the statistical uncertainties of the pairwise kSZ measurements.
This demonstrates that the simple scaling of eq.~\ref{eq:pksz} is indeed sufficient, and that internal cluster motions and/or correlations between velocity and optical depth do not have a strong impact on the pairwise kSZ signal. Furthermore, it shows that the $y$-$\tau$ scaling relation can be used to predict the correct optical depth to be used in eq.~\ref{eq:pksz}.
Overall, our findings demonstrate that future detections of the pairwise kSZ signal can indeed be used as a probe of cosmology.

\subsection{Observational prospects}
\label{subsec:observations}

In the above, we have shown that under idealised circumstances the pairwise kSZ signal combined with the $y$-$\tau$ scaling relation can yield an unbiased estimate of the mean pairwise velocity. 
Here we now discuss whether the inevitable non-idealities of real data affect this conclusion.

\subsubsection{Measurements from realistic CMB data}
For measuring the average Compton-$y$ parameter and the pairwise kSZ signal, we have used tSZ- and kSZ-only maps, which are of course not readily available from a real CMB experiment.
Thanks to its characteristic multi-frequency signature, however, the tSZ signal can be extracted from multi-frequency CMB data relatively easily:
one can either construct a $y$-map through appropriate linear combination of different bands (e.g.~\citealt{Hill2014,Planck2015_SZmap}), or combine measurements at multiple frequencies to estimate the tSZ signal of every cluster from a fit to the SED (e.g.~\citealt{Soergel2016}).
Therefore we do not expect any bias to the measurement of $y$ when real data including primary CMB, noise, and foregrounds are considered.
The kSZ signal, on the other hand, is lacking this distinctive frequency signature, so it cannot be extracted with multi-frequency cleaning techniques.
Therefore we have to rely on the differential structure of the pairwise estimator to remove all contributions that do not depend on cluster pair separation; this includes thermal SZ, primary CMB, instrumental noise, and foregrounds such as radio and infrared galaxies.

We have repeated our estimate of the pairwise kSZ signal from a combined kSZ+tSZ map at 150 GHz.
To preclude any biases from the large tSZ signal of high-mass clusters (e.g.~\citealt{Flender2015,Soergel2016}), here we remove all clusters with $M_{500c} > 10^{14}~\msol$ from the sample. Thanks to the steepness of the halo mass function and the pairwise nature of our signal, this has only a very small impact on the overall signal-to-noise ratio.
We are still able to reconstruct an unbiased estimate of the mean pairwise velocity from this combined kSZ+tSZ map, but the significance is roughly halved:
we find $a = \{1.01 \pm 0.09, 1.03 \pm 0.06, 0.97 \pm 0.06, 1.03 \pm 0.07, 1.09 \pm 0.09 \}$ in the five redshift slices. 
Other potential strategies for precluding bias from tSZ and mitigating this loss of signal-to-noise are creating a tSZ-free CMB map by appropriate linear combination, or alternatively working at 217 GHz where the tSZ does not contribute.
Which of these strategies is going to recover most of the signal depends on the detailed properties of both the CMB data and the cluster catalogue.

We have also computed the pairwise kSZ signal from a simulated 150 GHz map including
kSZ and a Gaussian realization of primary CMB, foregrounds, and instrumental noise.
In particular, we compute the CMB power spectrum with \texttt{CAMB} \citep{CAMB}, model radio and infrared galaxies according to the best-fitting power spectra from \cite{George2014}, and assume an instrumental noise level of $7~\mu\mathrm{K}\,\arcm$ (representative of AdvACTPol, \citealt{Calabrese2014}).
Again, we recover an unbiased estimate of the mean pairwise velocity, but at significantly reduced signal-to-noise: we find 
$S/N \simeq \{2, 4, 7, 3, 2\}$ for the five redshift bins.
The estimated S/N is, however, based on only $\simeq 1600~\mathrm{deg}^2$, or a sky coverage of $\simeq 4\%$. For an experiment with larger sky coverage, these numbers should be rescaled by a factor of $\simeq (f_\mathrm{sky}/0.04)^{1/2}$. In the case of AdvActPol $\times$ DESI \citep{DESI} we therefore expect an approximately three times higher signal-to-noise ratio.

It is worth noting that at these low noise levels the power on cluster scales is already becoming dominated by foregrounds (as opposed to instrumental noise for ACT and SPT-SZ).
Some of these foregrounds, most notably emission from radio and infrared galaxies, are correlated with the kSZ signal because a fraction of these galaxies resides in group- or cluster-sized haloes. At frequencies below 217 GHz (where the tSZ contribution is negative), they will additionally anti-correlate with the tSZ signal, so that the contributions from these contaminants mix and partially cancel in a non-trivial way.
Modelling them as Gaussian random fields with their measured power spectra is therefore clearly insufficient.
Instead, a population of radio- and infrared galaxies has to be extracted from the simulation (e.g.~\citealt{Dave2010,Fontanot2011,Hirschmann2014}).

\subsubsection{Other observational effects}
\label{sec:magn:otherobs}

\subparagraph{Cluster properties:}
We have assumed that the angular size of every cluster is known, which is of course not the case for real data. 
However, as argued above, a mass-observable relation provides an estimate for the average radius of clusters of a given mass and redshift. 
We have tested that binning clusters by their angular size and repeating the AP procedure with an aperture given by the binned radii does not change our results significantly.

\subparagraph{Miscentring:} 

A further potential source of systematic uncertainty are offsets between the cluster centre and the peak of the SZ signal, which affect both the average $y$ and the measured pairwise kSZ signal.	
If they are of astrophysical origin, as in the case of unrelaxed or merging clusters, their effects are captured in our hydrodynamical simulations and therefore in our scaling relation.
If, on the other hand, they are caused by imperfections in the cluster finding algorithm (such as misidentification of the central galaxy in optical cluster finders), they introduce an additional bias, which we estimate here.

The effect of miscentring can be expressed as a convolution of the true kSZ (tSZ) signal profile with a miscentring distribution (e.g.~\citealt{Johnston:2007uc,Saro2015}). Miscentring therefore broadens the average cluster profile, similarly to the instrumental beam (see Appendix \ref{app:beam_impact}).
For concreteness, we use the model calibrated by \citet{Saro2015} using DES clusters and SPT data. It describes the 2D offset distribution as a mixture of two normal distributions, a well-centred population ($p_0 = 0.63$, $\sigma_0 = 0.07 \, \theta_{500}$), and a population with worse centring ($p_1=0.37$, $\sigma_1 = 0.25 \, \theta_{500}$).
Using the model of Appendix \ref{app:beam_impact}, we estimate that this level of miscentring reduces the average optical depth by a few percent, broadly consistent with the findings of \protect\citet{Flender2015} and \protect\cite{Soergel2016}. 

We note that the observational calibration by \citet{Saro2015} captures both astrophysical offsets and potential misidentifications of the BCG. \citet{Gupta2016} calibrate the same model on hydrodynamical simulations and find similar results, suggesting that most of the miscentring is of astrophysical origin. As these offsets are already included in the calibration of our scaling relation, using the full miscentring model in our estimate is a conservative choice.
Also, while \cite{Saro2015} calibrate the model using mostly more massive clusters, \citet{Gupta2016} obtain comparable results and no strong trend in mass down to $M_{500c} \simeq 10^{14}~\msol$.
If, on the other hand, the observational centring properties are significantly worse for low-mass clusters or groups, miscentring would have a larger effect.\footnote{
See also \citet{Calafut2017} for a detailed observational study of the impact of miscentring on the pairwise kSZ signal.}

\subsection{Systematic uncertainties from sub-grid physics}
\label{subsec:systematics}

Our simulation includes prescriptions for the relevant feedback processes, which allow to properly reproduce various observational
properties of the ICM (see Sec.~\ref{subsec:sims}). 
There are, however, still a number of uncertainties associated with these sub-grid models and their implementation.
As extreme examples, various authors in the past compared ICM properties of simulations with and without AGN feedback
(e.g.~\citealt{Puchwein2008,2013MNRAS.431.1487P,2014MNRAS.438..195P,2017MNRAS.467.3827P}, B16). 
In particular, B16 repeated the calibration of the $y$-$\tau$ scaling relation
for a simulation run with and without AGN feedback, 
finding 6-8$\%$ difference at their default aperture of $1.3\arcmin$ and a decreasing trend towards larger apertures.
This should be taken as a very conservative upper bound for the systematic uncertainty on the $y$-$\tau$ relation.

Recent simulations for the first time match various observed ICM properties: they are able to reproduce cool-core and non-cool-core clusters \citep{2015ApJ...813L..17R,2017MNRAS.467.3827P}, as well as matching the observed chemical footprint 
\citep{2017MNRAS.468..531B,2017Galax...5...35D}. 
Changing sub-grid models without losing the current level of agreement between simulations and observations is no longer a simple task,
and is even more complicated for the large simulation boxes that we consider. Therefore we adopt the values of B16 as a conservative estimate
of the systematic uncertainty in the scaling relation.
In reality, however, these errors will likely be smaller considering variations in current simulations.
As already noted in B16, this systematic uncertainty is nonetheless larger than the statistical errors on the scaling relation.
The calibration of the $y$-$\tau$ relation on simulations is therefore currently limited by the unknown real uncertainty on the feedback prescriptions.

It is worth noting that currently the pairwise kSZ signal has only been detected at the $4\sigma$ level \citep{Soergel2016,deBernardis2016}.
This number is expected to increase significantly with data from the second- and third-generation receivers at ACT and SPT (e.g.~\citealt{Flender2015}), so that eventually the statistical uncertainties are going to become comparable to the current systematic uncertainty on the scaling relation.
However, we expect that future data and/or more advanced analyses going beyond the power spectrum are also going to help with further improving the feedback models in the simulations.
For example, measurements of the bispectrum of secondary CMB anisotropies (e.g.~\citealt{Crawford2014,Coulton2017}) improve our ability to separate the SZ signal from foregrounds, and might also provide additional constraints on feedback processes.
Combining future tSZ and kSZ data also enables improved constraints on cluster thermodynamics and feedback efficiencies \citep{Battaglia2017}.

Another possibility to reduce the dependency of the SZ signal on baryonic physics would be to excise the core of the clusters, similar to what is done in X-ray analyses. This would, however, require data with significantly higher resolution. This could be achieved from interferometric observations, which are, however, observationally much more expensive.

\subsection{Implications for other kSZ detection techniques}
In the above we have focused exclusively on the real-space pairwise estimator, as it is a robust technique that has already led to several kSZ detections from different experiments.
Here we now discuss the implications of our findings for other techniques.

The Fourier space version of the pairwise estimator derived by \cite{Sugiyama2016,Sugiyama2017} captures the effect of redshift-space distortions on the pairwise kSZ signal by decomposing the signal into multipoles, with the dipole approximately corresponding to the standard real-space measurement.
In principle, the different $k$-dependence of the dipole and octopole would allow to marginalize over the amplitude (which depends on the optical depth).
In practice, however, the signal-to-noise ratio is completely dominated by the dipole: \cite{Sugiyama2016} forecast $S/N \simeq 30$ for the dipole from a survey volume of $1~h^{-3}\mathrm{Gpc}^3$ and for CMB-S4 \citep{CMBS4_sciencebook} noise levels, compared to only $S/N \simeq 3$ for the octopole.
Therefore additional information on the amplitude from the $y$-$\tau$ relation would still be valuable to break the degeneracy between optical depth and velocity.

Velocity reconstruction-based techniques \citep{Li2014,Schaan2015,Planck_KSZ} suffer from a similar same degeneracy between velocity and optical depth (e.g.~\citealt{Alonso2016}), so also here the $y$-$\tau$ relation could be a useful tool.
The same holds true for estimates of the cluster velocity dispersion from the kSZ \citep{Planck_kSZ_dispersion}.

With future CMB observatories with a larger number of frequency bands, such as CCAT \citep{CCAT} or CMB-S4, it might also be possible to isolate the kSZ signal of a larger number of clusters than currently possible.
The degeneracy between optical depth and velocity can then be broken by including relativistic corrections to the SZ spectrum and/or X-ray-based temperature priors \citep{Sehgal2005,Mittal2017}. 
However, this approach is mostly limited to relatively massive clusters.

Approaches based on smooth fields instead of haloes, such as the `projected fields' technique by \citet{Hill2016,Ferraro2016}, are sensitive to the free electron fraction instead of the optical depth.
Potentially a similar approach (e.g.~calibrating a relation to the tSZ power spectrum or bispectrum) could be applied here, but further work is required to study the relation between the different SZ spectra in detail.

\section{Conclusions}
\label{sec:conclusions}
The pairwise kSZ signal has significant potential as a probe of cosmology. It could allow for an independent measurement of the parameter combination $Hf\sigma_8^2$, breaking degeneracies between these parameters from other probes of the growth of structure.
Realising this potential, however, requires disentangling the cluster optical depth (and all the associated uncertainties on baryonic physics) from the cosmologically interesting mean pairwise velocity.

One promising avenue for this is to calibrate a scaling relation between the Compton-$y$ parameter and the optical depth on hydrodynamical simulations.
Given a measurement of the average tSZ signal of a cluster sample, this relation can then be used to predict the average optical depth.
Provided that internal cluster motions and/or correlations between velocity and optical depth do not have a significant impact on the signal, we can then recover the mean pairwise velocity from the kSZ signal.

Here we have investigated whether this procedure returns an unbiased estimate of $\mathcal{v}_{12}$ at the precision required by the high signal-to-noise measurements expected from future surveys. 
For this purpose we have created SZ maps and cluster catalogues from the largest ($L=2688~\mpc$) box of the \textit{Magneticum} simulation suite.
Thanks to its large box size, for the first time we have been able to produce SZ maps from hydrodynamical simulations that cover a sky area comparable to current high-resolution CMB experiments.

We have measured the average Compton-$y$ parameter within the virial radius by applying an aperture photometry filter with adaptive size to the $y$-map.
For measuring the optical depth we have applied the same procedure to the kSZ map, and then correlate the AP-filtered values with the cluster line-of-sight velocities. This yields an estimate of the average optical depth $\bar{\tau}$ as a function of mass and redshift.
We have then fit a scaling relation between the two quantities, paying particular attention to any potential biases which could arise when applying that relation to SZ measurements from real CMB data.

We have finally computed the pairwise kSZ signal using the same AP estimates as before. From the average $\overline{y}$ in every redshift slice, we have predicted the corresponding average $\bar{\tau}$, and have used it to convert $T_\mathrm{pkSZ}(r)$ into $\mathcal{v}_{12}(r)$. These estimates are then compared to the `true' values computed from the halo catalogue.
We have found excellent agreement between the reconstructed and true mean pairwise velocity.
In other words, we have demonstrated that the pairwise kSZ signal yields an unbiased estimate
of the cosmologically interesting mean pairwise velocity, and that the degeneracy between velocity and optical depth can be overcome by using the $\tau$-$y$ scaling relation.
Therefore the pairwise kSZ effect can indeed be used as a probe of cosmology.

The largest systematic uncertainty associated with our scaling relation are the remaining variations between the sub-grid feedback models implemented in hydrodynamical simulations.
While obtaining realistic estimates for the errors in the calibration of the scaling relation based on large cosmological
volumes will be a major task for the next generation of simulation campaigns, we have adopted the 6-8$\%$ from B16 as a conservative estimate of such systematic uncertainty. 
Although this is well below the statistical uncertainty of current (and expected near-future) pairwise kSZ detections, eventually the measurements will reach this systematic floor. Therefore, further work on the calibration of feedback models is required to realize the full potential of the pairwise kSZ signal as a cosmological probe.

\section*{Acknowledgements}
The authors would like to thank Nick Battaglia, Lindsey Bleem, Jens Chluba, Samuel Flender, Nikhel Gupta, and Ewald Puchwein for helpful discussions,
and the anonymous referee for helpful suggestions.
BS acknowledges support from an Isaac Newton Studentship at the University of Cambridge and from the Science and Technology Facilities Council (STFC). AS is supported by the ERC-StG ``ClustersXCosmo'', grant agreement 716762.
KD acknowledges the support of the DFG Cluster of Excellence ``Origin and Structure of the
Universe'' and the Transregio programme TR33 ``The Dark Universe''.

The calculations were carried out at the Leibniz Supercomputer Center
(LRZ) under the project pr86re. We are especially grateful for
the support by M.~Petkova through the Computational Center
for Particle and Astrophysics (C2PAP) and the support by N.~Hammer at LRZ when carrying out the Box0 simulation within the Extreme Scale-Out
Phase on the new SuperMUC Haswell extension system. 

BS further thanks the developers of the \textit{Cython} programming language (\citealt{Cython}, \url{http://cython.org/}), which was used extensively during the work on this paper.
Furthermore, the use of the following software packages is acknowledged: 
\textit{Astropy}, a community-developed core Python package for Astronomy (\citealt{astropy}, \url{http://www.astropy.org/});
\textit{Flipper} (\citealt{Das2009}, \url{https://github.com/sudeepdas/flipper});
\textit{CosmoloPy} (\url{http://roban.github.io/CosmoloPy}).




\bibliographystyle{mnras}
\bibliography{ksz_magneticum}

\begin{thebibliography}{}
\makeatletter
\relax
\def\mn@urlcharsother{\let\do\@makeother \do\$\do\&\do\#\do\^\do\_\do\%\do\~}
\def\mn@doi{\begingroup\mn@urlcharsother \@ifnextchar [ {\mn@doi@}
  {\mn@doi@[]}}
\def\mn@doi@[#1]#2{\def\@tempa{#1}\ifx\@tempa\@empty \href
  {http://dx.doi.org/#2} {doi:#2}\else \href {http://dx.doi.org/#2} {#1}\fi
  \endgroup}
\def\mn@eprint#1#2{\mn@eprint@#1:#2::\@nil}
\def\mn@eprint@arXiv#1{\href {http://arxiv.org/abs/#1} {{\tt arXiv:#1}}}
\def\mn@eprint@dblp#1{\href {http://dblp.uni-trier.de/rec/bibtex/#1.xml}
  {dblp:#1}}
\def\mn@eprint@#1:#2:#3:#4\@nil{\def\@tempa {#1}\def\@tempb {#2}\def\@tempc
  {#3}\ifx \@tempc \@empty \let \@tempc \@tempb \let \@tempb \@tempa \fi \ifx
  \@tempb \@empty \def\@tempb {arXiv}\fi \@ifundefined
  {mn@eprint@\@tempb}{\@tempb:\@tempc}{\expandafter \expandafter \csname
  mn@eprint@\@tempb\endcsname \expandafter{\@tempc}}}

\bibitem[\protect\citeauthoryear{{Abazajian} et~al.,}{{Abazajian}
  et~al.}{2016}]{CMBS4_sciencebook}
{Abazajian} K.~N.,  et~al., 2016, preprint, \href
  {http://adsabs.harvard.edu/abs/2016arXiv161002743A} {} (\mn@eprint {arXiv}
  {1610.02743})

\bibitem[\protect\citeauthoryear{{Afshordi}, {Lin}, {Nagai}  \&
  {Sanderson}}{{Afshordi} et~al.}{2007}]{Afshordi07}
{Afshordi} N.,  {Lin} Y.-T.,  {Nagai} D.,   {Sanderson} A.~J.~R.,  2007,
  \mn@doi [\mnras] {10.1111/j.1365-2966.2007.11776.x}, \href
  {http://ukads.nottingham.ac.uk/abs/2007MNRAS.378..293A} {378, 293}

\bibitem[\protect\citeauthoryear{{Aghanim}, {G{\'o}rski}  \& {Puget}}{{Aghanim}
  et~al.}{2001}]{Aghanim2001}
{Aghanim} N.,  {G{\'o}rski} K.~M.,   {Puget} J.-L.,  2001, \mn@doi [\aap]
  {10.1051/0004-6361:20010659}, \href
  {http://adsabs.harvard.edu/abs/2001A%26A...374....1A} {374, 1}

\bibitem[\protect\citeauthoryear{{Ahn} et~al.,}{{Ahn} et~al.}{2012}]{Ahn2012}
{Ahn} C.~P.,  et~al., 2012, \mn@doi [\apjs] {10.1088/0067-0049/203/2/21}, \href
  {http://adsabs.harvard.edu/abs/2012ApJS..203...21A} {203, 21}

\bibitem[\protect\citeauthoryear{{Alonso}, {Louis}, {Bull}  \&
  {Ferreira}}{{Alonso} et~al.}{2016}]{Alonso2016}
{Alonso} D.,  {Louis} T.,  {Bull} P.,   {Ferreira} P.~G.,  2016, \mn@doi [\prd]
  {10.1103/PhysRevD.94.043522}, \href
  {http://adsabs.harvard.edu/abs/2016PhRvD..94d3522A} {94, 043522}

\bibitem[\protect\citeauthoryear{{Arnaud}, {Pratt}, {Piffaretti},
  {B{\"o}hringer}, {Croston}  \& {Pointecouteau}}{{Arnaud}
  et~al.}{2010}]{Arnaud2010}
{Arnaud} M.,  {Pratt} G.~W.,  {Piffaretti} R.,  {B{\"o}hringer} H.,  {Croston}
  J.~H.,   {Pointecouteau} E.,  2010, \mn@doi [\aap]
  {10.1051/0004-6361/200913416}, \href
  {http://adsabs.harvard.edu/abs/2010A%26A...517A..92A} {517, A92}

\bibitem[\protect\citeauthoryear{{Astropy Collaboration}}{{Astropy
  Collaboration}}{2013}]{astropy}
{Astropy Collaboration} 2013, \mn@doi [\aap] {10.1051/0004-6361/201322068},
  \href {http://adsabs.harvard.edu/abs/2013A%26A...558A..33A} {558, A33}

\bibitem[\protect\citeauthoryear{{Battaglia}}{{Battaglia}}{2016}]{Battaglia2016}
{Battaglia} N.,  2016, \mn@doi [\jcap] {10.1088/1475-7516/2016/08/058}, \href
  {http://adsabs.harvard.edu/abs/2016JCAP...08..058B} {8, 058}

\bibitem[\protect\citeauthoryear{{Battaglia}, {Bond}, {Pfrommer}  \&
  {Sievers}}{{Battaglia} et~al.}{2012}]{Battaglia2012}
{Battaglia} N.,  {Bond} J.~R.,  {Pfrommer} C.,   {Sievers} J.~L.,  2012,
  \mn@doi [\apj] {10.1088/0004-637X/758/2/74}, \href
  {http://adsabs.harvard.edu/abs/2012ApJ...758...74B} {758, 74}

\bibitem[\protect\citeauthoryear{{Battaglia}, {Ferraro}, {Schaan}  \&
  {Spergel}}{{Battaglia} et~al.}{2017}]{Battaglia2017}
{Battaglia} N.,  {Ferraro} S.,  {Schaan} E.,   {Spergel} D.~N.,  2017, \mn@doi
  [\jcap] {10.1088/1475-7516/2017/11/040}, \href
  {http://adsabs.harvard.edu/abs/2017JCAP...11..040B} {11, 040}

\bibitem[\protect\citeauthoryear{{Beck} et~al.,}{{Beck}
  et~al.}{2016}]{2016MNRAS.455.2110B}
{Beck} A.~M.,  et~al., 2016, \mn@doi [\mnras] {10.1093/mnras/stv2443}, \href
  {http://adsabs.harvard.edu/abs/2016MNRAS.455.2110B} {455, 2110}

\bibitem[\protect\citeauthoryear{Bhattacharya \& Kosowsky}{Bhattacharya \&
  Kosowsky}{2007}]{Bhattacharya2006}
Bhattacharya S.,  Kosowsky A.,  2007, \mn@doi [\apj] {10.1086/517523}, 659, L83

\bibitem[\protect\citeauthoryear{Bhattacharya \& Kosowsky}{Bhattacharya \&
  Kosowsky}{2008}]{Bhattacharya2007}
Bhattacharya S.,  Kosowsky A.,  2008, \mn@doi [Phys.Rev.]
  {10.1103/PhysRevD.77.083004}, D77, 083004

\bibitem[\protect\citeauthoryear{{Biffi} et~al.,}{{Biffi}
  et~al.}{2017}]{2017MNRAS.468..531B}
{Biffi} V.,  et~al., 2017, \mn@doi [\mnras] {10.1093/mnras/stx444}, \href
  {http://adsabs.harvard.edu/abs/2017MNRAS.468..531B} {468, 531}

\bibitem[\protect\citeauthoryear{Birkinshaw}{Birkinshaw}{1999}]{Birkinshaw1998}
Birkinshaw M.,  1999, \mn@doi [Phys.Rept.] {10.1016/S0370-1573(98)00080-5},
  310, 97

\bibitem[\protect\citeauthoryear{{Bleem} et~al.,}{{Bleem}
  et~al.}{2015}]{bleem15}
{Bleem} L.~E.,  et~al., 2015, \mn@doi [\apjs] {10.1088/0067-0049/216/2/27},
  \href {http://adsabs.harvard.edu/abs/2015ApJS..216...27B} {216, 27}

\bibitem[\protect\citeauthoryear{{Bocquet}, {Saro}, {Dolag}  \&
  {Mohr}}{{Bocquet} et~al.}{2016}]{Bocquet2016}
{Bocquet} S.,  {Saro} A.,  {Dolag} K.,   {Mohr} J.~J.,  2016, \mn@doi [\mnras]
  {10.1093/mnras/stv2657}, \href
  {http://adsabs.harvard.edu/abs/2016MNRAS.456.2361B} {456, 2361}

\bibitem[\protect\citeauthoryear{{Bryan} \& {Norman}}{{Bryan} \&
  {Norman}}{1998}]{BryanNorman98}
{Bryan} G.~L.,  {Norman} M.~L.,  1998, \mn@doi [\apj] {10.1086/305262}, \href
  {http://adsabs.harvard.edu/abs/1998ApJ...495...80B} {495, 80}

\bibitem[\protect\citeauthoryear{{Calabrese} et~al.,}{{Calabrese}
  et~al.}{2014}]{Calabrese2014}
{Calabrese} E.,  et~al., 2014, \mn@doi [\jcap] {10.1088/1475-7516/2014/08/010},
  \href {http://adsabs.harvard.edu/abs/2014JCAP...08..010C} {8, 010}

\bibitem[\protect\citeauthoryear{{Calafut}, {Bean}  \& {Yu}}{{Calafut}
  et~al.}{2017}]{Calafut2017}
{Calafut} V.,  {Bean} R.,   {Yu} B.,  2017, preprint, \href
  {http://adsabs.harvard.edu/abs/2017arXiv171001755C} {} (\mn@eprint {arXiv}
  {1710.01755})

\bibitem[\protect\citeauthoryear{Carlstrom, Holder  \& Reese}{Carlstrom
  et~al.}{2002}]{Carlstrom2002}
Carlstrom J.~E.,  Holder G.~P.,   Reese E.~D.,  2002, \mn@doi [Ann.Rev.\aap]
  {10.1146/annurev.astro.40.060401.093803}, 40, 643

\bibitem[\protect\citeauthoryear{{Chluba} \& {Mannheim}}{{Chluba} \&
  {Mannheim}}{2002}]{Chluba2002}
{Chluba} J.,  {Mannheim} K.,  2002, \mn@doi [\aap]
  {10.1051/0004-6361:20021429}, \href
  {http://adsabs.harvard.edu/abs/2002A%26A...396..419C} {396, 419}

\bibitem[\protect\citeauthoryear{{Cooray} \& {Chen}}{{Cooray} \&
  {Chen}}{2002}]{Cooray2002}
{Cooray} A.,  {Chen} X.,  2002, \mn@doi [\apj] {10.1086/340582}, \href
  {http://adsabs.harvard.edu/abs/2002ApJ...573...43C} {573, 43}

\bibitem[\protect\citeauthoryear{{Coulton} et~al.,}{{Coulton}
  et~al.}{2017}]{Coulton2017}
{Coulton} W.~R.,  et~al., 2017, preprint, \href
  {http://adsabs.harvard.edu/abs/2017arXiv171107879C} {} (\mn@eprint {arXiv}
  {1711.07879})

\bibitem[\protect\citeauthoryear{{Crawford} et~al.,}{{Crawford}
  et~al.}{2014}]{Crawford2014}
{Crawford} T.~M.,  et~al., 2014, \mn@doi [\apj] {10.1088/0004-637X/784/2/143},
  \href {http://adsabs.harvard.edu/abs/2014ApJ...784..143C} {784, 143}

\bibitem[\protect\citeauthoryear{Dalcin, Bradshaw, Smith, Citro, Behnel  \&
  Seljebotn}{Dalcin et~al.}{2010}]{Cython}
Dalcin L.,  Bradshaw R.,  Smith K.,  Citro C.,  Behnel S.,   Seljebotn D.~S.,
  2010, \mn@doi [Computing in Science & Engineering]
  {doi.ieeecomputersociety.org/10.1109/MCSE.2010.118}, 13, 31

\bibitem[\protect\citeauthoryear{{Dark Energy Survey Collaboration}}{{Dark
  Energy Survey Collaboration}}{2016}]{DESnonDEoverview}
{Dark Energy Survey Collaboration} 2016, \mn@doi [\mnras]
  {10.1093/mnras/stw641}, \href
  {http://adsabs.harvard.edu/abs/2016MNRAS.460.1270D} {460, 1270}

\bibitem[\protect\citeauthoryear{{Das}, {Hajian}  \& {Spergel}}{{Das}
  et~al.}{2009}]{Das2009}
{Das} S.,  {Hajian} A.,   {Spergel} D.~N.,  2009, \mn@doi [\prd]
  {10.1103/PhysRevD.79.083008}, \href
  {http://adsabs.harvard.edu/abs/2009PhRvD..79h3008D} {79, 083008}

\bibitem[\protect\citeauthoryear{{Dav{\'e}}, {Finlator}, {Oppenheimer},
  {Fardal}, {Katz}, {Kere{\v s}}  \& {Weinberg}}{{Dav{\'e}}
  et~al.}{2010}]{Dave2010}
{Dav{\'e}} R.,  {Finlator} K.,  {Oppenheimer} B.~D.,  {Fardal} M.,  {Katz} N.,
  {Kere{\v s}} D.,   {Weinberg} D.~H.,  2010, \mn@doi [\mnras]
  {10.1111/j.1365-2966.2010.16395.x}, \href
  {http://adsabs.harvard.edu/abs/2010MNRAS.404.1355D} {404, 1355}

\bibitem[\protect\citeauthoryear{{De Bernardis} et~al.,}{{De Bernardis}
  et~al.}{2017}]{deBernardis2016}
{De Bernardis} F.,  et~al., 2017, \mn@doi [\jcap]
  {10.1088/1475-7516/2017/03/008}, \href
  {http://adsabs.harvard.edu/abs/2017JCAP...03..008D} {3, 008}

\bibitem[\protect\citeauthoryear{{Diaferio}, {Sunyaev}  \& {Nusser}}{{Diaferio}
  et~al.}{2000}]{Diaferio2000}
{Diaferio} A.,  {Sunyaev} R.~A.,   {Nusser} A.,  2000, \mn@doi [\apjl]
  {10.1086/312627}, \href {http://adsabs.harvard.edu/abs/2000ApJ...533L..71D}
  {533, L71}

\bibitem[\protect\citeauthoryear{{Diaferio} et~al.,}{{Diaferio}
  et~al.}{2005}]{Diaferio2004}
{Diaferio} A.,  et~al., 2005, \mn@doi [\mnras]
  {10.1111/j.1365-2966.2004.08586.x}, \href
  {http://adsabs.harvard.edu/abs/2005MNRAS.356.1477D} {356, 1477}

\bibitem[\protect\citeauthoryear{{Diego} \& {Partridge}}{{Diego} \&
  {Partridge}}{2010}]{Diego10}
{Diego} J.~M.,  {Partridge} B.,  2010, \mn@doi [\mnras]
  {10.1111/j.1365-2966.2009.15949.x}, \href
  {http://ukads.nottingham.ac.uk/abs/2010MNRAS.402.1179D} {402, 1179}

\bibitem[\protect\citeauthoryear{{Dolag} \& {Sunyaev}}{{Dolag} \&
  {Sunyaev}}{2013}]{Dolag2013}
{Dolag} K.,  {Sunyaev} R.,  2013, \mn@doi [\mnras] {10.1093/mnras/stt579},
  \href {http://adsabs.harvard.edu/abs/2013MNRAS.432.1600D} {432, 1600}

\bibitem[\protect\citeauthoryear{{Dolag}, {Hansen}, {Roncarelli}  \&
  {Moscardini}}{{Dolag} et~al.}{2005}]{Dolag2005}
{Dolag} K.,  {Hansen} F.~K.,  {Roncarelli} M.,   {Moscardini} L.,  2005,
  \mn@doi [\mnras] {10.1111/j.1365-2966.2005.09452.x}, \href
  {http://adsabs.harvard.edu/abs/2005MNRAS.363...29D} {363, 29}

\bibitem[\protect\citeauthoryear{{Dolag}, {Borgani}, {Murante}  \&
  {Springel}}{{Dolag} et~al.}{2009}]{Dolag2009}
{Dolag} K.,  {Borgani} S.,  {Murante} G.,   {Springel} V.,  2009, \mn@doi
  [\mnras] {10.1111/j.1365-2966.2009.15034.x}, \href
  {http://adsabs.harvard.edu/abs/2009MNRAS.399..497D} {399, 497}

\bibitem[\protect\citeauthoryear{{Dolag}, {Komatsu}  \& {Sunyaev}}{{Dolag}
  et~al.}{2016}]{Dolag2015}
{Dolag} K.,  {Komatsu} E.,   {Sunyaev} R.,  2016, \mn@doi [\mnras]
  {10.1093/mnras/stw2035}, \href
  {http://adsabs.harvard.edu/abs/2016MNRAS.463.1797D} {463, 1797}

\bibitem[\protect\citeauthoryear{{Dolag}, {Mevius}  \& {Remus}}{{Dolag}
  et~al.}{2017}]{2017Galax...5...35D}
{Dolag} K.,  {Mevius} E.,   {Remus} R.-S.,  2017, \mn@doi [Galaxies]
  {10.3390/galaxies5030035}, \href
  {http://adsabs.harvard.edu/abs/2017Galax...5...35D} {5, 35}

\bibitem[\protect\citeauthoryear{Dunkley, Calabrese, Sievers, Addison,
  Battaglia  et~al.}{Dunkley et~al.}{2013}]{Dunkley2013}
Dunkley J.,  Calabrese E.,  Sievers J.,  Addison G.,  Battaglia N.,   et~al.,
  2013, \mn@doi [J. Cosmology Astropart. Phys.]
  {10.1088/1475-7516/2013/07/025}, 1307, 025

\bibitem[\protect\citeauthoryear{{Erler}, {Basu}, {Chluba}  \&
  {Bertoldi}}{{Erler} et~al.}{2017}]{Erler2017}
{Erler} J.,  {Basu} K.,  {Chluba} J.,   {Bertoldi} F.,  2017, preprint, \href
  {http://adsabs.harvard.edu/abs/2017arXiv170901187E} {} (\mn@eprint {arXiv}
  {1709.01187})

\bibitem[\protect\citeauthoryear{{Ferraro}, {Hill}, {Battaglia}, {Liu}  \&
  {Spergel}}{{Ferraro} et~al.}{2016}]{Ferraro2016}
{Ferraro} S.,  {Hill} J.~C.,  {Battaglia} N.,  {Liu} J.,   {Spergel} D.~N.,
  2016, \mn@doi [\prd] {10.1103/PhysRevD.94.123526}, \href
  {http://adsabs.harvard.edu/abs/2016PhRvD..94l3526F} {94, 123526}

\bibitem[\protect\citeauthoryear{Ferreira, Juszkiewicz, Feldman, Davis  \&
  Jaffe}{Ferreira et~al.}{1999}]{Ferreira1998}
Ferreira P.,  Juszkiewicz R.,  Feldman H.,  Davis M.,   Jaffe A.~H.,  1999,
  \mn@doi [\apj] {10.1086/311959}, 515, L1

\bibitem[\protect\citeauthoryear{{Flender}, {Bleem}, {Finkel}, {Habib},
  {Heitmann}  \& {Holder}}{{Flender} et~al.}{2016}]{Flender2015}
{Flender} S.,  {Bleem} L.,  {Finkel} H.,  {Habib} S.,  {Heitmann} K.,
  {Holder} G.,  2016, \mn@doi [\apj] {10.3847/0004-637X/823/2/98}, \href
  {http://adsabs.harvard.edu/abs/2016ApJ...823...98F} {823, 98}

\bibitem[\protect\citeauthoryear{{Flender}, {Nagai}  \& {McDonald}}{{Flender}
  et~al.}{2017}]{Flender2017}
{Flender} S.,  {Nagai} D.,   {McDonald} M.,  2017, \mn@doi [\apj]
  {10.3847/1538-4357/aa60bf}, \href
  {http://adsabs.harvard.edu/abs/2017ApJ...837..124F} {837, 124}

\bibitem[\protect\citeauthoryear{{Fontanot} \& {Somerville}}{{Fontanot} \&
  {Somerville}}{2011}]{Fontanot2011}
{Fontanot} F.,  {Somerville} R.~S.,  2011, \mn@doi [\mnras]
  {10.1111/j.1365-2966.2011.19245.x}, \href
  {http://adsabs.harvard.edu/abs/2011MNRAS.416.2962F} {416, 2962}

\bibitem[\protect\citeauthoryear{George, Reichardt, Aird, Benson, Bleem
  et~al.}{George et~al.}{2015}]{George2014}
George E.,  Reichardt C.,  Aird K.,  Benson B.,  Bleem L.,   et~al., 2015,
  \mn@doi [\apj] {10.1088/0004-637X/799/2/177}, 799, 177

\bibitem[\protect\citeauthoryear{{Gupta}, {Saro}, {Mohr}, {Dolag}  \&
  {Liu}}{{Gupta} et~al.}{2017}]{Gupta2016}
{Gupta} N.,  {Saro} A.,  {Mohr} J.~J.,  {Dolag} K.,   {Liu} J.,  2017, \mn@doi
  [\mnras] {10.1093/mnras/stx715}, \href
  {http://adsabs.harvard.edu/abs/2017MNRAS.469.3069G} {469, 3069}

\bibitem[\protect\citeauthoryear{Haehnelt \& Tegmark}{Haehnelt \&
  Tegmark}{1996}]{Haehnelt1996}
Haehnelt M.~G.,  Tegmark M.,  1996, \mn@doi [\mnras] {10.1093/mnras/279.2.545},
  279, 545

\bibitem[\protect\citeauthoryear{{Hajian}, {Battaglia}, {Spergel}, {Bond},
  {Pfrommer}  \& {Sievers}}{{Hajian} et~al.}{2013}]{Hajian13}
{Hajian} A.,  {Battaglia} N.,  {Spergel} D.~N.,  {Bond} J.~R.,  {Pfrommer} C.,
   {Sievers} J.~L.,  2013, \mn@doi [\jcap] {10.1088/1475-7516/2013/11/064},
  \href {http://ukads.nottingham.ac.uk/abs/2013JCAP...11..064H} {11, 064}

\bibitem[\protect\citeauthoryear{Hand, Addison, Aubourg, Battaglia, Battistelli
   et~al.}{Hand et~al.}{2012}]{Hand2012}
Hand N.,  Addison G.~E.,  Aubourg E.,  Battaglia N.,  Battistelli E.~S.,
  et~al., 2012, \mn@doi [Phys.Rev.Lett.] {10.1103/PhysRevLett.109.041101}, 109,
  041101

\bibitem[\protect\citeauthoryear{Hartlap, Simon  \& Schneider}{Hartlap
  et~al.}{2007}]{Hartlap2006}
Hartlap J.,  Simon P.,   Schneider P.,  2007, \mn@doi [\aap]
  {10.1051/0004-6361:20066170}, 464, 399

\bibitem[\protect\citeauthoryear{Hasselfield, Hilton, Marriage, Addison
  et~al.}{Hasselfield et~al.}{2013}]{Hasselfield2013}
Hasselfield M.,  Hilton M.,  Marriage T.~A.,  Addison G.~E.,   et~al., 2013,
  \mn@doi [J. Cosmology Astropart. Phys.] {10.1088/1475-7516/2013/07/008},
  1307, 008

\bibitem[\protect\citeauthoryear{{Hern{\'a}ndez-Monteagudo}, {Ma}, {Kitaura},
  {Wang}, {G{\'e}nova-Santos}, {Mac{\'{\i}}as-P{\'e}rez}  \&
  {Herranz}}{{Hern{\'a}ndez-Monteagudo} et~al.}{2015}]{HM2015}
{Hern{\'a}ndez-Monteagudo} C.,  {Ma} Y.-Z.,  {Kitaura} F.~S.,  {Wang} W.,
  {G{\'e}nova-Santos} R.,  {Mac{\'{\i}}as-P{\'e}rez} J.,   {Herranz} D.,  2015,
  \mn@doi [Physical Review Letters] {10.1103/PhysRevLett.115.191301}, \href
  {http://adsabs.harvard.edu/abs/2015PhRvL.115s1301H} {115, 191301}

\bibitem[\protect\citeauthoryear{{Hill} \& {Spergel}}{{Hill} \&
  {Spergel}}{2014}]{Hill2014}
{Hill} J.~C.,  {Spergel} D.~N.,  2014, \mn@doi [\jcap]
  {10.1088/1475-7516/2014/02/030}, \href
  {http://adsabs.harvard.edu/abs/2014JCAP...02..030H} {2, 030}

\bibitem[\protect\citeauthoryear{{Hill}, {Ferraro}, {Battaglia}, {Liu}  \&
  {Spergel}}{{Hill} et~al.}{2016}]{Hill2016}
{Hill} J.~C.,  {Ferraro} S.,  {Battaglia} N.,  {Liu} J.,   {Spergel} D.~N.,
  2016, \mn@doi [Physical Review Letters] {10.1103/PhysRevLett.117.051301},
  \href {http://adsabs.harvard.edu/abs/2016PhRvL.117e1301H} {117, 051301}

\bibitem[\protect\citeauthoryear{{Hirschmann}, {Dolag}, {Saro}, {Bachmann},
  {Borgani}  \& {Burkert}}{{Hirschmann} et~al.}{2014}]{Hirschmann2014}
{Hirschmann} M.,  {Dolag} K.,  {Saro} A.,  {Bachmann} L.,  {Borgani} S.,
  {Burkert} A.,  2014, \mn@doi [\mnras] {10.1093/mnras/stu1023}, \href
  {http://adsabs.harvard.edu/abs/2014MNRAS.442.2304H} {442, 2304}

\bibitem[\protect\citeauthoryear{{Jaffe} \& {Kamionkowski}}{{Jaffe} \&
  {Kamionkowski}}{1998}]{Jaffe1998}
{Jaffe} A.~H.,  {Kamionkowski} M.,  1998, \mn@doi [\prd]
  {10.1103/PhysRevD.58.043001}, \href
  {http://adsabs.harvard.edu/abs/1998PhRvD..58d3001J} {58, 043001}

\bibitem[\protect\citeauthoryear{{Johnston} et~al.,}{{Johnston}
  et~al.}{2007}]{Johnston:2007uc}
{Johnston} D.~E.,  et~al., 2007, preprint, \href
  {http://adsabs.harvard.edu/abs/2007arXiv0709.1159J} {} (\mn@eprint {arXiv}
  {0709.1159})

\bibitem[\protect\citeauthoryear{{Kaiser}}{{Kaiser}}{1986}]{Kaiser1986}
{Kaiser} N.,  1986, \mn@doi [\mnras] {10.1093/mnras/222.2.323}, \href
  {http://adsabs.harvard.edu/abs/1986MNRAS.222..323K} {222, 323}

\bibitem[\protect\citeauthoryear{Keisler \& Schmidt}{Keisler \&
  Schmidt}{2013}]{Keisler2012}
Keisler R.,  Schmidt F.,  2013, \mn@doi [\apj] {10.1088/2041-8205/765/2/L32},
  765, L32

\bibitem[\protect\citeauthoryear{{Komatsu} et~al.,}{{Komatsu}
  et~al.}{2011}]{Komatsu2011}
{Komatsu} E.,  et~al., 2011, \mn@doi [\apjs] {10.1088/0067-0049/192/2/18},
  \href {http://adsabs.harvard.edu/abs/2011ApJS..192...18K} {192, 18}

\bibitem[\protect\citeauthoryear{{Levi} et~al.,}{{Levi} et~al.}{2013}]{DESI}
{Levi} M.,  et~al., 2013, preprint, \href
  {http://adsabs.harvard.edu/abs/2013arXiv1308.0847L} {} (\mn@eprint {arXiv}
  {1308.0847})

\bibitem[\protect\citeauthoryear{{Lewis}, {Challinor}  \& {Lasenby}}{{Lewis}
  et~al.}{2000}]{CAMB}
{Lewis} A.,  {Challinor} A.,   {Lasenby} A.,  2000, \mn@doi [\apj]
  {10.1086/309179}, \href {http://adsabs.harvard.edu/abs/2000ApJ...538..473L}
  {538, 473}

\bibitem[\protect\citeauthoryear{{Li}, {Angulo}, {White}  \& {Jasche}}{{Li}
  et~al.}{2014}]{Li2014}
{Li} M.,  {Angulo} R.~E.,  {White} S.~D.~M.,   {Jasche} J.,  2014, \mn@doi
  [\mnras] {10.1093/mnras/stu1224}, \href
  {http://adsabs.harvard.edu/abs/2014MNRAS.443.2311L} {443, 2311}

\bibitem[\protect\citeauthoryear{{Li}, {Ma}, {Remazeilles}  \& {Moodley}}{{Li}
  et~al.}{2017}]{LiYC2017}
{Li} Y.-C.,  {Ma} Y.-Z.,  {Remazeilles} M.,   {Moodley} K.,  2017, preprint,
  \href {http://adsabs.harvard.edu/abs/2017arXiv171010876L} {} (\mn@eprint
  {arXiv} {1710.10876})

\bibitem[\protect\citeauthoryear{Lueker, Reichardt, Schaffer, Zahn, Ade
  et~al.}{Lueker et~al.}{2010}]{lueker10}
Lueker M.,  Reichardt C.,  Schaffer K.,  Zahn O.,  Ade P.,   et~al., 2010,
  \mn@doi [\apj] {10.1088/0004-637X/719/2/1045}, 719, 1045

\bibitem[\protect\citeauthoryear{{Ma}, {Gong}, {Sui}  \& {He}}{{Ma}
  et~al.}{2017}]{Ma2017}
{Ma} Y.-Z.,  {Gong} G.-D.,  {Sui} N.,   {He} P.,  2017, preprint, \href
  {http://adsabs.harvard.edu/abs/2017arXiv171108756M} {} (\mn@eprint {arXiv}
  {1711.08756})

\bibitem[\protect\citeauthoryear{{McCarthy}, {Le Brun}, {Schaye}  \&
  {Holder}}{{McCarthy} et~al.}{2014}]{McCarthy2014}
{McCarthy} I.~G.,  {Le Brun} A.~M.~C.,  {Schaye} J.,   {Holder} G.~P.,  2014,
  \mn@doi [\mnras] {10.1093/mnras/stu543}, \href
  {http://adsabs.harvard.edu/abs/2014MNRAS.440.3645M} {440, 3645}

\bibitem[\protect\citeauthoryear{{Melchior} et~al.,}{{Melchior}
  et~al.}{2017}]{Melchior2016}
{Melchior} P.,  et~al., 2017, \mn@doi [\mnras] {10.1093/mnras/stx1053}, \href
  {http://adsabs.harvard.edu/abs/2017MNRAS.469.4899M} {469, 4899}

\bibitem[\protect\citeauthoryear{{Mittal}, {de Bernardis}  \&
  {Niemack}}{{Mittal} et~al.}{2017}]{Mittal2017}
{Mittal} A.,  {de Bernardis} F.,   {Niemack} M.~D.,  2017, preprint, \href
  {http://adsabs.harvard.edu/abs/2017arXiv170806365M} {} (\mn@eprint {arXiv}
  {1708.06365})

\bibitem[\protect\citeauthoryear{{Mueller}, {de Bernardis}, {Bean}  \&
  {Niemack}}{{Mueller} et~al.}{2015a}]{Mueller2014b}
{Mueller} E.-M.,  {de Bernardis} F.,  {Bean} R.,   {Niemack} M.~D.,  2015a,
  \mn@doi [\prd] {10.1103/PhysRevD.92.063501}, \href
  {http://adsabs.harvard.edu/abs/2015PhRvD..92f3501M} {92, 063501}

\bibitem[\protect\citeauthoryear{{Mueller}, {de Bernardis}, {Bean}  \&
  {Niemack}}{{Mueller} et~al.}{2015b}]{Mueller2014a}
{Mueller} E.-M.,  {de Bernardis} F.,  {Bean} R.,   {Niemack} M.~D.,  2015b,
  \mn@doi [\apj] {10.1088/0004-637X/808/1/47}, \href
  {http://adsabs.harvard.edu/abs/2015ApJ...808...47M} {808, 47}

\bibitem[\protect\citeauthoryear{{Nagai}, {Kravtsov}  \& {Kosowsky}}{{Nagai}
  et~al.}{2003}]{Nagai2003}
{Nagai} D.,  {Kravtsov} A.~V.,   {Kosowsky} A.,  2003, \mn@doi [\apj]
  {10.1086/368281}, \href {http://adsabs.harvard.edu/abs/2003ApJ...587..524N}
  {587, 524}

\bibitem[\protect\citeauthoryear{{Nelson}, {Lau}  \& {Nagai}}{{Nelson}
  et~al.}{2014}]{Nelson2014}
{Nelson} K.,  {Lau} E.~T.,   {Nagai} D.,  2014, \mn@doi [\apj]
  {10.1088/0004-637X/792/1/25}, \href
  {http://adsabs.harvard.edu/abs/2014ApJ...792...25N} {792, 25}

\bibitem[\protect\citeauthoryear{{Park}, {Komatsu}, {Shapiro}, {Koda}  \&
  {Mao}}{{Park} et~al.}{2016}]{Park2015}
{Park} H.,  {Komatsu} E.,  {Shapiro} P.~R.,  {Koda} J.,   {Mao} Y.,  2016,
  \mn@doi [\apj] {10.3847/0004-637X/818/1/37}, \href
  {http://adsabs.harvard.edu/abs/2016ApJ...818...37P} {818, 37}

\bibitem[\protect\citeauthoryear{{Park}, {Alvarez}  \& {Bond}}{{Park}
  et~al.}{2017}]{Park2017}
{Park} H.,  {Alvarez} M.~A.,   {Bond} J.~R.,  2017, preprint, \href
  {http://adsabs.harvard.edu/abs/2017arXiv171002792P} {} (\mn@eprint {arXiv}
  {1710.02792})

\bibitem[\protect\citeauthoryear{{Plagge} et~al.,}{{Plagge}
  et~al.}{2010}]{plagge10}
{Plagge} T.,  et~al., 2010, \mn@doi [\apj] {10.1088/0004-637X/716/2/1118},
  \href {http://adsabs.harvard.edu/abs/2010ApJ...716.1118P} {716, 1118}

\bibitem[\protect\citeauthoryear{{Planck Collaboration}}{{Planck
  Collaboration}}{2011}]{Bartlett11}
{Planck Collaboration} 2011, \mn@doi [\aap] {10.1051/0004-6361/201116489},
  \href {http://ukads.nottingham.ac.uk/abs/2011A%26A...536A..12P} {536, A12}

\bibitem[\protect\citeauthoryear{{Planck Collaboration}}{{Planck
  Collaboration}}{2013a}]{Planck12}
{Planck Collaboration} 2013a, \mn@doi [\aap] {10.1051/0004-6361/201220040},
  \href {http://adsabs.harvard.edu/abs/2013A%26A...550A.131P} {550, A131}

\bibitem[\protect\citeauthoryear{{Planck Collaboration}}{{Planck
  Collaboration}}{2013b}]{Planckint_hotgas}
{Planck Collaboration} 2013b, \mn@doi [\aap] {10.1051/0004-6361/201220941},
  \href {http://adsabs.harvard.edu/abs/2013A%26A...557A..52P} {557, A52}

\bibitem[\protect\citeauthoryear{{Planck Collaboration}}{{Planck
  Collaboration}}{2016a}]{Planck_KSZ}
{Planck Collaboration} 2016a, \mn@doi [\aap] {10.1051/0004-6361/201526328},
  \href {http://adsabs.harvard.edu/abs/2016A%26A...586A.140P} {586, A140}

\bibitem[\protect\citeauthoryear{{Planck Collaboration}}{{Planck
  Collaboration}}{2016b}]{Planck2015params}
{Planck Collaboration} 2016b, \mn@doi [\aap] {10.1051/0004-6361/201525830},
  \href {http://adsabs.harvard.edu/abs/2016A%26A...594A..13P} {594, A13}

\bibitem[\protect\citeauthoryear{{Planck Collaboration}}{{Planck
  Collaboration}}{2016c}]{Planck2015_SZmap}
{Planck Collaboration} 2016c, \mn@doi [\aap] {10.1051/0004-6361/201525826},
  \href {http://adsabs.harvard.edu/abs/2016A%26A...594A..22P} {594, A22}

\bibitem[\protect\citeauthoryear{{Planck Collaboration}}{{Planck
  Collaboration}}{2016d}]{Planck2015_SZ}
{Planck Collaboration} 2016d, \mn@doi [\aap] {10.1051/0004-6361/201525823},
  \href {http://adsabs.harvard.edu/abs/2016A%26A...594A..27P} {594, A27}

\bibitem[\protect\citeauthoryear{{Planck Collaboration}}{{Planck
  Collaboration}}{2017}]{Planck_kSZ_dispersion}
{Planck Collaboration} 2017, preprint, \href
  {http://adsabs.harvard.edu/abs/2017arXiv170700132P} {} (\mn@eprint {arXiv}
  {1707.00132})

\bibitem[\protect\citeauthoryear{{Planelles}, {Borgani}, {Dolag}, {Ettori},
  {Fabjan}, {Murante}  \& {Tornatore}}{{Planelles}
  et~al.}{2013}]{2013MNRAS.431.1487P}
{Planelles} S.,  {Borgani} S.,  {Dolag} K.,  {Ettori} S.,  {Fabjan} D.,
  {Murante} G.,   {Tornatore} L.,  2013, \mn@doi [\mnras]
  {10.1093/mnras/stt265}, \href
  {http://adsabs.harvard.edu/abs/2013MNRAS.431.1487P} {431, 1487}

\bibitem[\protect\citeauthoryear{{Planelles}, {Borgani}, {Fabjan}, {Killedar},
  {Murante}, {Granato}, {Ragone-Figueroa}  \& {Dolag}}{{Planelles}
  et~al.}{2014}]{2014MNRAS.438..195P}
{Planelles} S.,  {Borgani} S.,  {Fabjan} D.,  {Killedar} M.,  {Murante} G.,
  {Granato} G.~L.,  {Ragone-Figueroa} C.,   {Dolag} K.,  2014, \mn@doi [\mnras]
  {10.1093/mnras/stt2141}, \href
  {http://adsabs.harvard.edu/abs/2014MNRAS.438..195P} {438, 195}

\bibitem[\protect\citeauthoryear{{Planelles} et~al.,}{{Planelles}
  et~al.}{2017}]{2017MNRAS.467.3827P}
{Planelles} S.,  et~al., 2017, \mn@doi [\mnras] {10.1093/mnras/stx318}, \href
  {http://adsabs.harvard.edu/abs/2017MNRAS.467.3827P} {467, 3827}

\bibitem[\protect\citeauthoryear{{Pollina}, {Hamaus}, {Dolag}, {Weller},
  {Baldi}  \& {Moscardini}}{{Pollina} et~al.}{2017}]{Pollina2017}
{Pollina} G.,  {Hamaus} N.,  {Dolag} K.,  {Weller} J.,  {Baldi} M.,
  {Moscardini} L.,  2017, \mn@doi [\mnras] {10.1093/mnras/stx785}, \href
  {http://adsabs.harvard.edu/abs/2017MNRAS.469..787P} {469, 787}

\bibitem[\protect\citeauthoryear{{Puchwein}, {Sijacki}  \&
  {Springel}}{{Puchwein} et~al.}{2008}]{Puchwein2008}
{Puchwein} E.,  {Sijacki} D.,   {Springel} V.,  2008, \mn@doi [\apjl]
  {10.1086/593352}, \href {http://adsabs.harvard.edu/abs/2008ApJ...687L..53P}
  {687, L53}

\bibitem[\protect\citeauthoryear{{Rasia} et~al.,}{{Rasia}
  et~al.}{2015}]{2015ApJ...813L..17R}
{Rasia} E.,  et~al., 2015, \mn@doi [\apjl] {10.1088/2041-8205/813/1/L17}, \href
  {http://adsabs.harvard.edu/abs/2015ApJ...813L..17R} {813, L17}

\bibitem[\protect\citeauthoryear{{Remus}, {Dolag}  \& {Hoffmann}}{{Remus}
  et~al.}{2017a}]{Remus2017b}
{Remus} R.-S.,  {Dolag} K.,   {Hoffmann} T.,  2017a, \mn@doi [Galaxies]
  {10.3390/galaxies5030049}, \href
  {http://adsabs.harvard.edu/abs/2017Galax...5...49R} {5, 49}

\bibitem[\protect\citeauthoryear{{Remus}, {Dolag}, {Naab}, {Burkert},
  {Hirschmann}, {Hoffmann}  \& {Johansson}}{{Remus} et~al.}{2017b}]{Remus2017}
{Remus} R.-S.,  {Dolag} K.,  {Naab} T.,  {Burkert} A.,  {Hirschmann} M.,
  {Hoffmann} T.~L.,   {Johansson} P.~H.,  2017b, \mn@doi [\mnras]
  {10.1093/mnras/stw2594}, \href
  {http://adsabs.harvard.edu/abs/2017MNRAS.464.3742R} {464, 3742}

\bibitem[\protect\citeauthoryear{{Rephaeli} \& {Lahav}}{{Rephaeli} \&
  {Lahav}}{1991}]{Rephaeli1991}
{Rephaeli} Y.,  {Lahav} O.,  1991, \mn@doi [\apj] {10.1086/169950}, \href
  {http://adsabs.harvard.edu/abs/1991ApJ...372...21R} {372, 21}

\bibitem[\protect\citeauthoryear{{Saro} et~al.,}{{Saro}
  et~al.}{2014}]{Saro2014}
{Saro} A.,  et~al., 2014, \mn@doi [\mnras] {10.1093/mnras/stu575}, \href
  {http://adsabs.harvard.edu/abs/2014MNRAS.440.2610S} {440, 2610}

\bibitem[\protect\citeauthoryear{{Saro} et~al.,}{{Saro}
  et~al.}{2015}]{Saro2015}
{Saro} A.,  et~al., 2015, \mn@doi [\mnras] {10.1093/mnras/stv2141}, \href
  {http://adsabs.harvard.edu/abs/2015MNRAS.454.2305S} {454, 2305}

\bibitem[\protect\citeauthoryear{{Saro} et~al.,}{{Saro}
  et~al.}{2017}]{Saro2017}
{Saro} A.,  et~al., 2017, \mn@doi [\mnras] {10.1093/mnras/stx594}, \href
  {http://adsabs.harvard.edu/abs/2017MNRAS.468.3347S} {468, 3347}

\bibitem[\protect\citeauthoryear{Sayers, Mroczkowski, Zemcov, Korngut, Bock
  et~al.}{Sayers et~al.}{2013}]{Sayers2013}
Sayers J.,  Mroczkowski T.,  Zemcov M.,  Korngut P.,  Bock J.,   et~al., 2013,
  \mn@doi [\apj] {10.1088/0004-637X/778/1/52}, 778, 52

\bibitem[\protect\citeauthoryear{{Schaan} et~al.,}{{Schaan}
  et~al.}{2016}]{Schaan2015}
{Schaan} E.,  et~al., 2016, \mn@doi [\prd] {10.1103/PhysRevD.93.082002}, \href
  {http://adsabs.harvard.edu/abs/2016PhRvD..93h2002S} {93, 082002}

\bibitem[\protect\citeauthoryear{{Sehgal}, {Kosowsky}  \& {Holder}}{{Sehgal}
  et~al.}{2005}]{Sehgal2005}
{Sehgal} N.,  {Kosowsky} A.,   {Holder} G.,  2005, \mn@doi [\apj]
  {10.1086/497258}, \href {http://adsabs.harvard.edu/abs/2005ApJ...635...22S}
  {635, 22}

\bibitem[\protect\citeauthoryear{Sheth, Diaferio, Hui  \& Scoccimarro}{Sheth
  et~al.}{2001}]{Sheth2000}
Sheth R.~K.,  Diaferio A.,  Hui L.,   Scoccimarro R.,  2001, \mn@doi [\mnras]
  {10.1046/j.1365-8711.2001.04457.x}, 326, 463

\bibitem[\protect\citeauthoryear{{Simet}, {McClintock}, {Mandelbaum}, {Rozo},
  {Rykoff}, {Sheldon}  \& {Wechsler}}{{Simet} et~al.}{2017}]{Simet2016}
{Simet} M.,  {McClintock} T.,  {Mandelbaum} R.,  {Rozo} E.,  {Rykoff} E.,
  {Sheldon} E.,   {Wechsler} R.~H.,  2017, \mn@doi [\mnras]
  {10.1093/mnras/stw3250}, \href
  {http://adsabs.harvard.edu/abs/2017MNRAS.466.3103S} {466, 3103}

\bibitem[\protect\citeauthoryear{{Soergel} et~al.,}{{Soergel}
  et~al.}{2016}]{Soergel2016}
{Soergel} B.,  et~al., 2016, \mn@doi [\mnras] {10.1093/mnras/stw1455}, \href
  {http://adsabs.harvard.edu/abs/2016MNRAS.461.3172S} {461, 3172}

\bibitem[\protect\citeauthoryear{{Soergel}, {Giannantonio}, {Efstathiou},
  {Puchwein}  \& {Sijacki}}{{Soergel} et~al.}{2017}]{Soergel2017}
{Soergel} B.,  {Giannantonio} T.,  {Efstathiou} G.,  {Puchwein} E.,   {Sijacki}
  D.,  2017, \mn@doi [\mnras] {10.1093/mnras/stx492}, \href
  {http://adsabs.harvard.edu/abs/2017MNRAS.468..577S} {468, 577}

\bibitem[\protect\citeauthoryear{{Springel}}{{Springel}}{2005}]{Springel2005}
{Springel} V.,  2005, \mn@doi [\mnras] {10.1111/j.1365-2966.2005.09655.x},
  \href {http://adsabs.harvard.edu/abs/2005MNRAS.364.1105S} {364, 1105}

\bibitem[\protect\citeauthoryear{{Springel}, {White}, {Tormen}  \&
  {Kauffmann}}{{Springel} et~al.}{2001}]{Springel2001}
{Springel} V.,  {White} S.~D.~M.,  {Tormen} G.,   {Kauffmann} G.,  2001,
  \mn@doi [\mnras] {10.1046/j.1365-8711.2001.04912.x}, \href
  {http://adsabs.harvard.edu/abs/2001MNRAS.328..726S} {328, 726}

\bibitem[\protect\citeauthoryear{{Steinborn}, {Dolag}, {Comerford},
  {Hirschmann}, {Remus}  \& {Teklu}}{{Steinborn} et~al.}{2016}]{Steinborn2016}
{Steinborn} L.~K.,  {Dolag} K.,  {Comerford} J.~M.,  {Hirschmann} M.,  {Remus}
  R.-S.,   {Teklu} A.~F.,  2016, \mn@doi [\mnras] {10.1093/mnras/stw316}, \href
  {http://adsabs.harvard.edu/abs/2016MNRAS.458.1013S} {458, 1013}

\bibitem[\protect\citeauthoryear{{Sugiyama}, {Okumura}  \&
  {Spergel}}{{Sugiyama} et~al.}{2017a}]{Sugiyama2017}
{Sugiyama} N.~S.,  {Okumura} T.,   {Spergel} D.~N.,  2017a, preprint, \href
  {http://adsabs.harvard.edu/abs/2017arXiv170507449S} {} (\mn@eprint {arXiv}
  {1705.07449})

\bibitem[\protect\citeauthoryear{{Sugiyama}, {Okumura}  \&
  {Spergel}}{{Sugiyama} et~al.}{2017b}]{Sugiyama2016}
{Sugiyama} N.~S.,  {Okumura} T.,   {Spergel} D.~N.,  2017b, \mn@doi [\jcap]
  {10.1088/1475-7516/2017/01/057}, \href
  {http://adsabs.harvard.edu/abs/2017JCAP...01..057S} {1, 057}

\bibitem[\protect\citeauthoryear{{Sunyaev} \& {Zeldovich}}{{Sunyaev} \&
  {Zeldovich}}{1970}]{SZ1970}
{Sunyaev} R.~A.,  {Zeldovich} Y.~B.,  1970, \mn@doi [\apss]
  {10.1007/BF00653471}, \href
  {http://adsabs.harvard.edu/abs/1970Ap%26SS...7....3S} {7, 3}

\bibitem[\protect\citeauthoryear{{Sunyaev} \& {Zeldovich}}{{Sunyaev} \&
  {Zeldovich}}{1972}]{SZ1972}
{Sunyaev} R.~A.,  {Zeldovich} Y.~B.,  1972, Comments on Astrophysics and Space
  Physics, \href {http://adsabs.harvard.edu/abs/1972CoASP...4..173S} {4, 173}

\bibitem[\protect\citeauthoryear{{Sunyaev} \& {Zeldovich}}{{Sunyaev} \&
  {Zeldovich}}{1980}]{SZ1980}
{Sunyaev} R.~A.,  {Zeldovich} I.~B.,  1980, \mnras, \href
  {http://adsabs.harvard.edu/abs/1980MNRAS.190..413S} {190, 413}

\bibitem[\protect\citeauthoryear{Swetz, Ade, Amiri, Appel, Battistelli
  et~al.}{Swetz et~al.}{2011}]{ACT}
Swetz D.,  Ade P.,  Amiri M.,  Appel J.,  Battistelli E.,   et~al., 2011,
  \mn@doi [\apjs] {10.1088/0067-0049/194/2/41}, 194, 41

\bibitem[\protect\citeauthoryear{{Teklu}, {Remus}, {Dolag}, {Beck}, {Burkert},
  {Schmidt}, {Schulze}  \& {Steinborn}}{{Teklu} et~al.}{2015}]{Teklu2015}
{Teklu} A.~F.,  {Remus} R.-S.,  {Dolag} K.,  {Beck} A.~M.,  {Burkert} A.,
  {Schmidt} A.~S.,  {Schulze} F.,   {Steinborn} L.~K.,  2015, \mn@doi [\apj]
  {10.1088/0004-637X/812/1/29}, \href
  {http://adsabs.harvard.edu/abs/2015ApJ...812...29T} {812, 29}

\bibitem[\protect\citeauthoryear{{Teklu}, {Remus}, {Dolag}  \&
  {Burkert}}{{Teklu} et~al.}{2017}]{Teklu2017}
{Teklu} A.~F.,  {Remus} R.-S.,  {Dolag} K.,   {Burkert} A.,  2017, \mn@doi
  [\mnras] {10.1093/mnras/stx2303}, \href
  {http://adsabs.harvard.edu/abs/2017MNRAS.472.4769T} {472, 4769}

\bibitem[\protect\citeauthoryear{{The Dark Energy Survey Collaboration}}{{The
  Dark Energy Survey Collaboration}}{2005}]{DES}
{The Dark Energy Survey Collaboration} 2005, preprint, \href
  {http://adsabs.harvard.edu/abs/2005astro.ph.10346T} {} (\mn@eprint {arXiv}
  {astro-ph/0510346})

\bibitem[\protect\citeauthoryear{{Woody} et~al.,}{{Woody} et~al.}{2012}]{CCAT}
{Woody} D.,  et~al., 2012, in Ground-based and Airborne Telescopes IV. p.
  84442M, \mn@doi{10.1117/12.925229}

\makeatother
\end{thebibliography}



\appendix

\section{Impact of the instrumental beam}
\label{app:beam_impact}
In this appendix we discuss the impact of the instrumental beam on the AP estimates of the optical depth.
Intuitively, the instrumental beam broadens the cluster profile, and therefore moves part of the signal from the inner AP disk to the outer ring.
As the latter is used for the background subtraction, the beam broadening leads to part of the signal being subtracted as well.
Here we develop a simplistic model to estimate the magnitude of this effect.

If the projected electron density profile of a cluster is given by $n_e(\mb{\theta})$, its average optical depth after beam-convolution and AP filtering scales as \citep{Sugiyama2017}
\beq
\langle \tau \rangle_A \propto \int \frac{d^2 \bm{\ell}}{(2\pi)^2} \tilde{W}_{AP}(\ell \theta_A) n(\bm{\ell}) B(\bm{\ell}) \,,
\label{eq:beamtest}
\eeq
where $n(\bm{\ell})$ and $B(\bm{\ell})$ are the Fourier transforms of the projected electron density and beam profile, respectively.
The relative impact of the instrumental beam on the estimate is then given by the ratio 
$\langle \tau \rangle_A/\langle \tau \rangle_A^\mathrm{no-beam}$,
where for $\langle \tau \rangle_A^\mathrm{no-beam}$ we simply set $B(\bm{\ell}) = 1$ in eq.~\ref{eq:beamtest}.

For simplicity we assume that $n_e(\mb{\theta})$ is given by a Gaussian profile with $\sigma = 0.5 \times \theta_\mathrm{vir}$.
This choice predicts that around $15\%$ of the gas mass are outside $\theta_\mathrm{vir}$, which is broadly in agreement with our findings in~Appendix \ref{app:tau_mgas}.
As in our main analysis, we use an aperture of $\theta_A = \theta_\mathrm{vir}$, and 
assume an instrumental beam with $\mathrm{FWHM} = 1.2\arcmin$ (representative of high-resolution CMB experiments like ACT and SPT).
We note that in this simplistic picture the result only depends on the ratio between beam FWHM and $\theta_\mathrm{vir}$.

Using $\theta_\mathrm{vir} = 2\arcmin$, which corresponds to the median projected virial radius of our central redshift bin, we find 
\beq
\langle \tau \rangle_A/\langle \tau \rangle_A^\mathrm{no-beam} \simeq 0.85 \ .
\eeq
If we use $\theta_\mathrm{vir} = 1\arcmin~(4\arcmin)$ instead, the resulting ratios are 0.52 and 0.96, respectively.
We therefore see that depending on the angular size of the cluster (and therefore the aperture) the impact of the beam on the AP estimate can be significant.

In~Appendix~\ref{app:tau_mgas} below we convert our estimates of $\langle \tau \rangle_A$ into a gas mass and compare to the `true' value measured from the simulation particle data. As part of this comparison, we also quantify the impact of the beam on the AP estimates directly from the simulations, see Fig.~\ref{fig:tau_mgas}.
Here the beam convolution reduces the AP estimate by $11\%$ ($5\%$) in the lowest (highest) mass bin with $\theta_\mathrm{vir} \simeq 2\arcmin$ ($4\arcmin$), broadly in agreement with the simplistic model we derived above.

When estimating the average Compton-$y$ parameter within an aperture we simply replace $n_e(\theta)$ in eq.~\ref{eq:beamtest} by $n_e(\theta)T(\theta)$.
Crucially, the angular dependences of tSZ and kSZ are different: the tSZ signal is more peaked towards the centre. 
Therefore, the relative effect of the beam on the two components as calculated above is different.
This could therefore have an impact on the slope of the $\tau$-$y$ relation.

In our main analysis we have convolved the SZ maps with an instrumental beam with $\mathrm{FWHM} = 1.2\arcmin$ before estimating the scaling relation.
Repeating the calculation above with $\mathrm{FWHM} = 1\arcmin$ and $1.4\arcmin$, we have tested that a small mismatch between the assumed and real beam size only affects the results at the level of a few percent, which is below the systematic uncertainty given by the sub-grid feedback models (see Sec.~\ref{subsec:systematics}).

We finally note that the same formalism can also be applied to estimate the impact of miscentring (see Sec.~\ref{sec:magn:otherobs}).
In this case, an extra factor $f_\mathrm{miscent}(\bm{\ell})$ is included in the integral of Eq.~\ref{eq:beamtest}, where $f_\mathrm{miscent}(\bm{\ell})$ is the Fourier transform of the miscentring distribution. The ratio
$\langle \tau \rangle_A/\langle \tau \rangle_A^\mathrm{no-miscent}$ then gives the relative impact of miscentring, where as before we set $f_\mathrm{miscent}(\bm{\ell}) = 1$ when computing $\langle \tau \rangle_A^\mathrm{no-miscent}$.

\section{Relation of the optical depth to the true gas mass}
\label{app:tau_mgas}
In the main analysis our discussion of the optical depth has focussed on using it in the context of pairwise kSZ measurements.  
Here we now investigate if the average optical depth can also be used as a probe of the true gas mass of the clusters.

The average optical depth within an aperture $\theta_A$ relates to the \textit{cylindrical} gas mass as
\begin{align}
\begin{split}
M_\mathrm{gas,A}^\mathrm{cyl} 
&= \int_\mathrm{cyl} \, \dint V \rho_\mathrm{gas} 
= \frac{d_A^2 \mu_e m_p}{\sigma_T} \int_{\theta_A} \dint \Omega \int \dint l \, n_e \sigma_T \\
&= \frac{\mu_e m_p}{\sigma_T} \pi \left(\theta_A d_A\right)^2 \langle \tau \rangle_A \, ,
\label{eq:mgas_app}
\end{split}
\end{align}
where $\mu_e \simeq 1.14$ is the mean particle weight per electron (assuming primordial abundances) and $m_p$ is the proton mass.
This relation provides us with a way of comparing our measurement of the optical depth from the main analysis to the `true' gas mass in the simulations.

However, computing $M_\mathrm{gas}^\mathrm{cyl}$ for all clusters in our large simulation box and for various redshifts and apertures would be computationally unfeasible.
Therefore we perform this comparison using the second-largest \textit{Magneticum} simulation box. 
This `Box1' has a sidelength of $L = 896~\mpc$ and the same resolution as the larger `Box0' that we used for the main analysis.
Cluster catalogues and SZ maps from this box were produced in the same way as for the large box (albeit with thinner redshift slices), and have already been presented in \cite{Saro2014,Dolag2015,Gupta2016}.

To preclude any bias from the beam convolution (see Appendix~\ref{app:beam_impact}), we perform this comparison using the unconvolved maps.
Furthermore, we perform this comparison using a representatively chosen subset of clusters (15 mass bins and 5 redshift bins, 100 randomly chosen clusters per bin).
For these objects, we calculate $M_\mathrm{gas}^\mathrm{cyl}$ for various apertures directly from the simulation particle data, including particles with up to $10~\mpc$ line-of-sight distance from the cluster centre.
We then compare this direct integration to an estimate of the gas mass obtained from our measurements of $\overline{\tau}$ via eq.~\ref{eq:mgas_app}, both for the average and the AP filter.

We note that the AP filter subtracts contributions in an annulus of $\theta_A < \theta < \sqrt{2} \theta_A$ from the signal within a disk with $\theta < \theta_A$. As clusters are not perfect circles in projection and do not have a sharp edge at any given radius, the AP filter will by construction yield an estimate of the gas mass that is biased low.
The level of bias will be largest for small apertures ($\theta_A \lesssim \theta_{500}$) and should decrease when including more of the cluster outskirts ($\theta_A \gtrsim \theta_\mathrm{vir}$).
On the other hand, the gas mass derived from the average-filtered $\overline{\tau}$ should be a much better estimator of the true cylindrical gas mass.

\begin{figure}
	\includegraphics[width=\columnwidth]{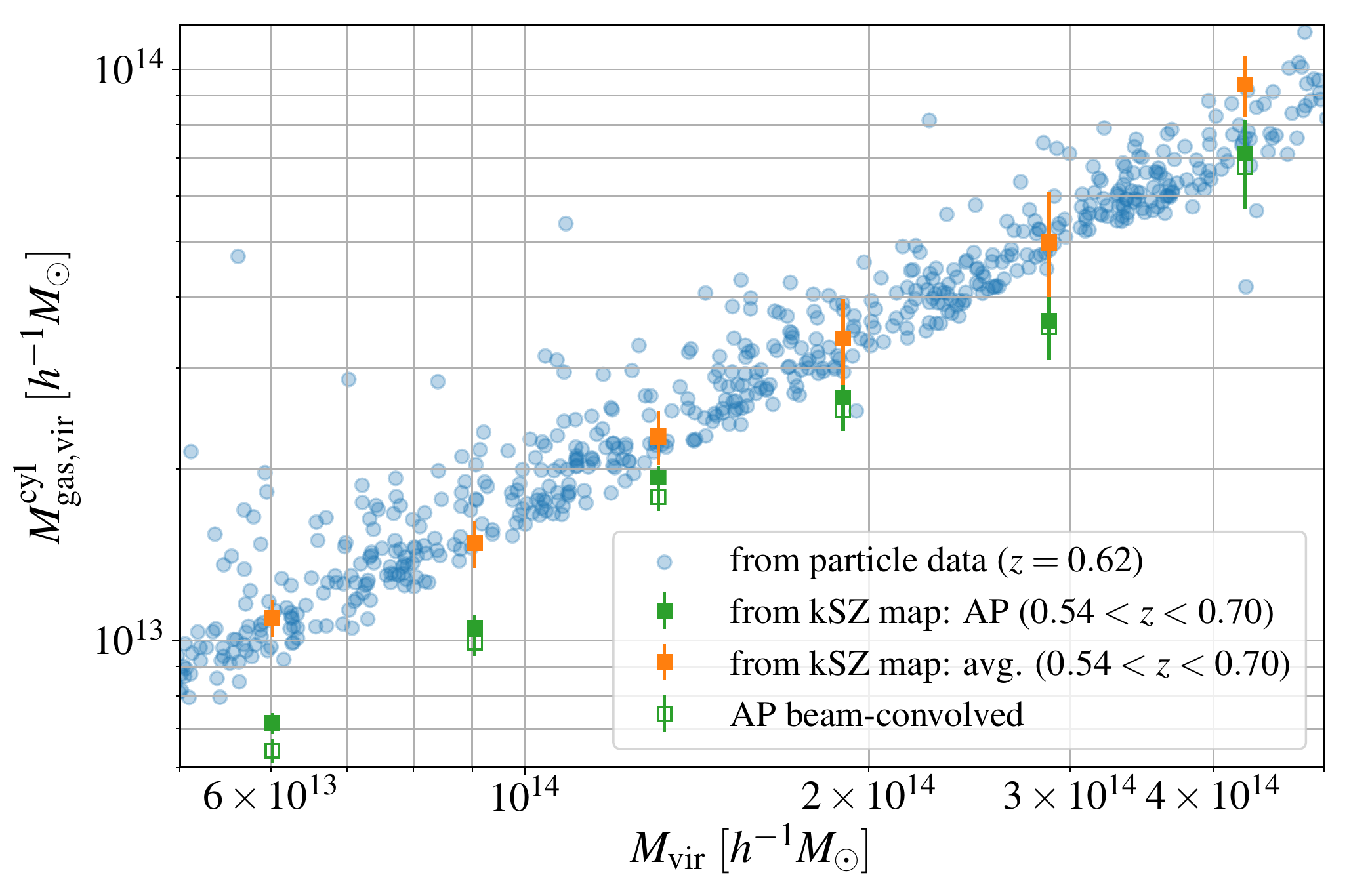}
	\caption{Estimates of the gas mass from the averaged optical depth: We compare here $M_\mathrm{gas}^\mathrm{cyl}$ as estimated from the average- and AP-filtered kSZ maps to the `true' value obtained directly from the simulation particle data. While the main comparison is performed without convolving with the instrumental beam, 
	we also show the effect of including the beam for the AP filter (see Appendix~\ref{app:beam_impact}).}
	\label{fig:tau_mgas}
\end{figure}

We show in Fig.~\ref{fig:tau_mgas} an example of the comparison for $z \simeq 0.6$ and $\theta_A = \theta_\mathrm{vir}$.
Here we indeed observe that the AP-estimate of $M_\mathrm{gas}^\mathrm{cyl}$ is  $\simeq 25\%$ lower than the true value, while the average filter yields the correct estimate of $M_\mathrm{gas}^\mathrm{cyl}$.
We have repeated this test for various apertures between $\theta_{500}$ and $1.5 \times \theta_\mathrm{vir}$ and obtained the expected trend of decreasing bias of the AP estimate at larger apertures.
Furthermore, we have repeated these comparisons for the other redshift bins and have obtained comparable results.

These comparisons also demonstrate that the way of measuring $\overline{\tau}$ that we describe in Sec.~\ref{subsec:meas_tau_y} above indeed yields the expected results.

\bsp	
\label{lastpage}
\end{document}